\documentclass[aps,twocolumn,preprintnumbers,amsmath,amssymb,floatfix,superscriptaddress]{revtex4}
\usepackage{graphicx}
\usepackage{color}
\def\ExB{{\bf E}\times{\bf B}}

\def\tHp{\tau_{\rm Hp}}

\def\SHp{S_{\rm Hp}}
\def\RHp{Re_{\rm Hp}}

\def\qRes{q_{\rm s}}

\def\rsi{r_{{\rm s}i}}
\def\rsj{r_{{\rm s}j}}
\def\rsa{r_{\rm s1}}
\def\rsb{r_{\rm s2}}
\def\rsc{r_{\rm s3}}

\def\mPeak{m_{\rm peak}}
\def\gPeak{\gamma_{\rm peak}}
\def\gDrive{\gamma_{\rm drive}}
\def\gMax{\gamma_{\rm max}}

\def\flPsi{\widetilde{\psi}}

\def\eqPhi{\overline{\phi}}
\def\eqVor{\overline{u}}
\def\eqCur{\overline{j}}

\def\gLin{\gamma_{\rm lin}}
\def\gKin{\gamma^{\rm kin}}
\def\gMag{\gamma^{\rm mag}}
\def\EKin{E^{\rm kin}}
\def\EMag{E^{\rm mag}}

\def\Ma{M^{(1)}}
\def\Mb{M^{(2)}}
\def\Mc{M^{(3)}}

\definecolor{gray}{rgb}{0.5,0.5,0.5}%
\definecolor{dred}{rgb}{0.5,0.0,0.0}%

\begin{document}

\title{Nonlinear evolution of the $m=1$ internal kink mode in the
presence of magnetohydrodynamic turbulence}

\author{Andreas~Bierwage}
\email{abierwag@uci.edu}
\affiliation{Graduate School of Energy Science, Kyoto University, Gokasho, Uji, Kyoto 611-0011, Japan}
\altaffiliation[Present address: ]{Department of Physics and Astronomy, University of California, Irvine, CA 92697, U.S.A.}
\author{Sadruddin~Benkadda}
\email{benkadda@up.univ-mrs.fr}
\affiliation{Equipe Dynamique des Syst\`{e}mes Complexes, UMR 6633 CNRS-Universit\'{e} de Provence, 13397 Marseille, France}
\author{Satoshi~Hamaguchi}
\email{hamaguch@ppl.eng.osaka-u.ac.jp}
\affiliation{Center for Atomic and Molecular Technologies, Osaka University, 2-1 Yamadaoka, Suita, Osaka 565-0871, Japan}
\author{Masahiro~Wakatani}
\thanks{deceased}
\affiliation{Graduate School of Energy Science, Kyoto University, Gokasho, Uji, Kyoto 611-0011, Japan}

\date{\today}

\begin{abstract}
The nonlinear evolution of the $m=1$ internal kink mode is studied numerically in a setting where the tokamak core plasma is surrounded by a turbulent region with low magnetic shear. As a starting point we choose configurations with three nearby $q=1$ surfaces where triple tearing modes (TTMs) with high poloidal mode numbers $m$ are unstable. While the amplitudes are still small, the fast growing high-$m$ TTMs enhance the growth of the $m=1$ instability. This is interpreted as a fast sawtooth trigger mechanism. The TTMs lead to a partial collapse, leaving behind a turbulent belt with $q \simeq 1$ around the unreconnected core plasma. Although, full reconnection can occur if the core displacement grows large enough, it is shown that the turbulence may actively prevent further reconnection. This is qualitatively similar to experimentally observed partial sawtooth crashes with post-cursor oscillations due to a saturated internal kink.
\end{abstract}

\maketitle

\thispagestyle{empty}

%===========================================
\section{Introduction}

Abrupt ejection of thermal energy and particles from a magnetized high-temperature plasma is frequently observed in astrophysical and laboratory plasmas. In the case of tokamak experiments, internal disruption events known as sawteeth bring about a sudden collapse of the core temperature through an internal kink instability. A thorough understanding of the underlying physical processes is important for the efficient operation of a tokamak device and the prospective application of the tokamak concept for thermonuclear fusion reactors.

The aim of this paper is to demonstrate the effect of magnetohydrodynamic (MHD) turbulence
%(i.e., modes with high poloidal mode numbers $m$)
on the evolution of the $m=1$ internal kink mode. Motivated by recent results on the instability of current-driven high-$m$ multiple tearing modes \cite{Bierwage05a, Bierwage05b} we choose configurations with three $q=1$ resonant surfaces located a small distance apart. Here, $q$ is the tokamak safety factor (measuring the field line pitch) and its central value is taken to be well below unity ($q_0 < 1$). The plasma is taken to have a finite resistivity to enable magnetic reconnection. This system is, in addition to the resistive $m=1$ internal kink mode, unstable to a broad spectrum of triple tearing modes (TTMs) with helicity $q_{\rm res} = m/n = 1$. At present, we neglect two-fluid and kinetic effects and also ignore the roles of finite pressure and toroidal curvature.

In this setting we address some open questions with regard to internal disruptions in tokamaks \cite{Kuvshinov90, Migliuolo93, Hastie98, Porcelli04}. These include the issue of the rapid onset of a sawtooth crash, known as the trigger problem, and the possibility of partial reconnection and compound sawtooth crashes (e.g., \cite{Soltwisch86, Edwards86, Levinton93}). The fast sawtooth trigger is defined as a sudden transition from slow growth or stability to rapid growth of the $m=1$ mode (e.g.,~\cite{Aydemir92, WangBhatt93}). The partial collapse is defined as a sawtooth crash during which the central core region remains intact, so that $q_0$ remains below unity (e.g.,~\cite{WangBhatt95, BiskampSato97}). It is clear that explanations for the sawtooth trigger and partial reconnection events require nonlinear effects. Several possible mechanisms have been proposed in the past (see, e.g., Ref.~\cite{Hastie98} for a review). Of particular interest here are scenarios that consider dynamics related to plasma turbulence.
% in a broad sense, i.e., the role of perturbations that lie in a range of high mode %numbers and vary in time.

In regimes where the $m=1$ mode amplitude is still small, high-$m$ modes were previously shown to affect the growth of the $m=1$ instability. Micro-turbulence \cite{Sundaram80, Sen81} (oscillating modes) and TTMs \cite{Bierwage05a} (purely growing modes) were found to enhance the $m=1$ growth rate. In other related studies it was shown that the $m=1$ mode can be stabilized by local oscillations at the resonant surface \cite{Thyagaraja92, Thyagaraja93} and micro-turbulence in the region of the thin current layer may lead to enhanced effective resistivity and thus faster reconnection \cite{Matthaeus86, Drake94, Kim01, Aparicio98}. Conversely, viscosity may be increased, which reduces the reconnection rate \cite{Meiss82}. In a broader sense, magnetic braiding and field line stochasticity may also be viewed as ``turbulent'' dynamics, and these were also shown to yield enhanced growth rates for reconnecting modes \cite{Kaw79, Lazarian99}. With regard to the long-time nonlinear evolution, pressure-driven instabilities (e.g., ballooning) were shown to be able to lead to a saturation of the internal kink at finite amplitude (partial collapse) \cite{Dubois80, Bussac87, Nishimura99}.

The scenario considered here is comparatively simple and includes only a minimum of physical effects. In the first part of this paper the case of a TTM-driven internal kink mode \cite{Bierwage05a} is examined and two new results are presented: establishment of a fast growing $m=1$ mode structure at low amplitudes and, in regimes with high viscosity ($Pr \gtrsim 10$), explosive growth during the transition to the turbulent regime.

The transition to turbulence occurs via a partial (annular) collapse, whereupon a turbulent belt forms around the central core. Indeed, in Ref.~\cite{Porcelli96} the conjecture was made that partial sawtooth crashes may be associated with wide-spread MHD turbulence in the reconnected region. In Ref.~\cite{Bierwage05a} the possibility for the $m=1$ mode to saturate in such a state was demonstrated using a numerical simulation where the core was constrained to a linear motion in the poloidal plane.

The second part of this paper deals with the long-term evolution of the internal kink, while surrounded by a low-shear region with $q \simeq 1$, governed by MHD turbulence. The scenario considered here is more generic than that in Ref.~\cite{Bierwage05a} since we apply a fully random perturbation and allow each Fourier mode to alter its poloidal phase angle through interaction with other modes. The resulting changes in the kink flow lead to an irregular or ``meandering'' motion of the core. One case is presented where full reconnection occurs eventually. In another case, the kink is found to saturate and decay, which shows that the MHD turbulence may prevent full reconnection.

A reduced model is used for the simulations in order to obtain first insights on a fundamental level and to lay the foundations for further investigations with more realistic models. Physical effects ignored in this work, such as two-fluid, curvature, and finite-beta effects, may play a significant role. In future work it would be interesting to see under which conditions and in which way these effects alter certain quantitative and qualitative features of the results presented here, including the linear and nonlinear instability growth, the dominant mode numbers in the inter-resonance region, and the evolution of the internal kink.

This paper is organized as follows. The physical model is introduced in Sec.~\ref{sec:model} and the numerical method is described in Sec.~\ref{sec:numerics}. The equilibrium used, its linear instability characteristics and the initial perturbation applied are given in Sec.~\ref{sec:equlib-lin}. In Sec.~\ref{sec:kink} we present results on the early evolution of the $m=1$ mode in the presence of fast growing high-$m$ TTMs. The transition to turbulence is treated in Sec.~\ref{sec:trans} and the long-term evolution is described in Sec.~\ref{sec:recon}. A summary, further discussions and conclusions are given in Section~\ref{sec:conclusion}.

%===========================================
\section{Model}
\label{sec:model}

We use the reduced set of magnetohydrodynamic (RMHD) equations in a cylindrical geometry in the limit of zero beta \cite{Strauss76, NishikawaWakatani}. This model has proven to be useful in studies of MHD instabilities when the focus is on a qualitative description of fundamental aspects of the magnetized plasma system, as is the case here. The RMHD model governs the evolution of the magnetic flux function $\psi$ and the electrostatic potential $\phi$, as described in Ref.~\cite{Bierwage05b}. The normalized RMHD equations are
\begin{eqnarray}
\partial_t\psi &=& \left[\psi,\phi\right] - \partial_\zeta\phi - \SHp^{-1}\left(\hat{\eta}j - E_0\right)
\label{eq:rmhd1}
\\
\partial_t u &=& \left[u,\phi\right] + \left[j,\psi\right] + \partial_\zeta j + \RHp^{-1}\nabla_\perp^2 u.
\label{eq:rmhd2}
\end{eqnarray}

\noindent The time is measured in units of the poloidal Alfv\'{e}n time $\tHp = \sqrt{\mu_0 \rho_{\rm m}} a/B_0$ and the radial coordinate is normalized by the minor radius $a$ of the plasma. Here, $\rho_{\rm m}$ is the mass density and $B_0$ the strong axial magnetic field. The current density $j$ and the vorticity $u$ are related to $\psi$ and $\phi$ through $j = -\nabla_\perp^2\psi$ and $u = \nabla_\perp^2\phi$, respectively.

In order to provide a simple mechanism for magnetic reconnection, a resistive diffusion term is included in Eq.~(\ref{eq:rmhd1}). Its strength is measured by the magnetic Reynolds number $\SHp = \tau_\eta / \tHp$, with $\tau_\eta = a^2\mu_0/\eta_0$ being the resistive diffusion time and $\eta_0=\eta(r=0)$ the electrical resistivity in the plasma core. We use $\SHp = 10^6$, which is numerically efficient and physically reasonable in the framework of the model used.
%\cite{Vlad89, BiskampSato97, Aydemir97}.
Flow damping is provided by an ion viscosity term in Eq.~(\ref{eq:rmhd2}). Viscous dissipation is measured by the kinematic Reynolds number $\RHp = a^2/\nu\tHp$, where $\nu$ is the ion viscosity. Long-time calculations are performed for $\RHp = 10^6$ and $10^8$, as will be specified case by case.

In order to ensure that, in the absence of magnetic reconnection, the equilibrium remains unchanged, the source term $\SHp^{-1}E_0$ is included in Eq.~(\ref{eq:rmhd1}). With $E_0 = \hat{\eta}\eqCur$ it compensates the resistive dissipation of the equilibrium current. The loop voltage measured by $E_0$ is taken to be constant, so the resistivity profile is given in terms of the equilibrium current density distribution as $\hat{\eta}(r) = \eqCur(r=0)/\eqCur(r)$. For simplicity, the temporal variation of the resistivity profile $\hat{\eta}$ is neglected.

As in Ref.~\cite{Bierwage05b}, each field variable $f$ is decomposed into an equilibrium part $\overline{f}$ and a perturbation $\widetilde{f}$ as
\begin{equation}
f(r,\vartheta,\zeta,t) = \overline{f}(r) + \widetilde{f}(r, \vartheta, \zeta, t).
\end{equation}

\noindent The system is described in terms of the Fourier modes, $\psi_{m,n}$ and $\phi_{m,n}$, obtained from the expansion
\begin{equation}
f(r, \vartheta, \zeta, t) = \frac{1}{2}\sum_{m,n} f_{m,n}(r,t).e^{i(m\vartheta - n\zeta)} + {\rm c.c.},
\end{equation}

\noindent with $m$ being the poloidal mode number and $n$ the toroidal mode number. In the following, the $(m,n)$ subscripts will often be omitted for convenience. We consider only the dynamics within a given helicity $h=m/n={\rm const}$, so the problem is reduced to two dimensions.

For the description of the dynamics in this system, it is useful to define the helical flux function $\psi_*$ with corresponding current density $j_*$ as
\begin{equation}
\psi_* = \psi + \frac{n}{2m}r^2 \quad {\rm and} \quad j_* = -\nabla_\perp^2\psi_* = j - \frac{2n}{m}.
\label{eq:def_psih}
\end{equation}

\noindent The evolution of the individual Fourier modes is described in terms of their kinetic and magnetic energies,
\begin{equation}
\EKin_{m,n} = |\nabla\phi_{m,n}|^2 \quad {\rm and} \quad \EMag_{m,n} = |\nabla\psi_{m,n}|^2,
\label{eq:def-E}
\end{equation}

\noindent and the corresponding nonlinear growth rates,
\begin{equation}
\gKin_{m,n}(t) = \frac{{\rm d}\ln \EKin_{m,n}}{2{\rm d}t} \quad {\rm and} \quad \gMag_{m,n}(t) = \frac{{\rm d}\ln \EMag_{m,n}}{2{\rm d}t}.
\label{eq:def-gr}
\end{equation}

\noindent (these are amplitude growth rates, hence the factor $1/2$). In Eq.~(\ref{eq:def-E}), $|f_{m,n}|^2 \equiv \int\limits_0^1{\rm d}r.r.C_m |f_{m,n}(r)|^2$, with $C_{m=0} = 4\pi$ and $C_{m\neq 0} = 2\pi$.

%===========================================
\section{Numerical method}
\label{sec:numerics}

For the numerical solution of the model equations (\ref{eq:rmhd1}) and (\ref{eq:rmhd2}), a two-step predictor-corrector method is applied. In the first time step the dissipation terms are treated implicitly, all others explicitly, and the field variables are estimated at an intermediate time step $t+\Delta t/2$. The second is a full time step, $t \rightarrow t+\Delta t$, with the right-hand sides of Eqs.~(\ref{eq:rmhd1}) and (\ref{eq:rmhd2}) evaluated at the intermediate time step $t+\Delta t/2$ estimated before. In the nonlinear regime the time step size is of the order $\Delta t \sim 10^{-3}$.

128 Fourier modes (including $m=0$) are carried. The Poisson brackets $[f,g] = \frac{1}{r}(\partial_r f \partial_\vartheta g - \partial_r g \partial_\vartheta f)$ are evaluated in real space. This pseudo-spectral method has been applied together with an appropriate dealiasing technique. The outcomes of the long-term evolution in both cases studied, full reconnection and kink saturation, were also confirmed using 256 Fourier modes.

The radial coordinate is discretized using a non-uniformly spaced grid, with a grid density of up to $N_r^{-1} \approx 1/2000$ [$1/5000$ for numerical checks] in regions where sharp current density peaks occur. A fourth-order centered-finite-difference method is applied for the $\partial_r$-terms in the Poisson brackets. The Laplacians $\nabla_{\perp(m,n)}^2 = \frac{1}{r}\partial_r r\partial_r - m^2/r^2$ are evaluated at second-order accuracy (tridiagonal matrix equations).

Periodic boundary conditions are applied in the azimuthal and axial directions. At $r=1$ an ideally conducting wall is assumed, requiring all perturbations to be identical to zero at that location: $\widetilde{f}(r=1) = 0$ (fixed boundary, no vacuum region). At $r=0$, extraneous boundary conditions are applied to ensure smoothness: $\partial_r \widetilde{f}_{m=0}(r=0) = 0$ and $\widetilde{f}_{m \neq 0}(r=0) = 0$.

%===========================================
\section{Equilibrium, linear instability and initial perturbation}
\label{sec:equlib-lin}

\begin{figure}
[tbp]
\centering
\includegraphics[height=6.0cm,width=8.0cm]% aspect ratio: 1.333
{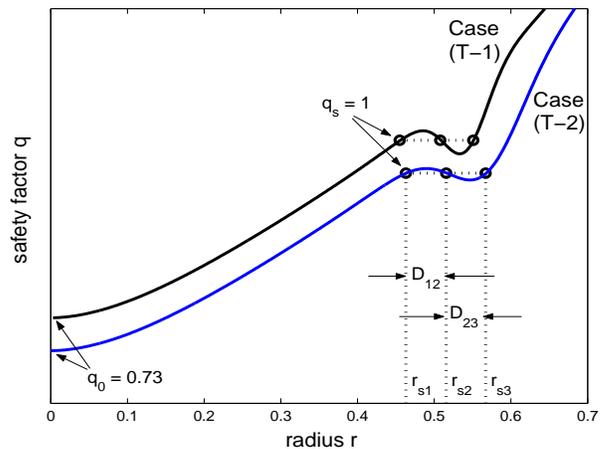}
\caption{(Color online). Safety factor profiles $q(r)$. Resonant surfaces are indicated by circles. Model parameters for both cases, labeled (T-1) and (T-2), are given in Table~\protect\ref{tab:equlib_3tm_q-parm}. The profile properties are listed in Table~\protect\ref{tab:equlib_3tm} and the dispersion relations are shown in Figs.~\protect\ref{fig:spec_3tm_T-1} and \protect\ref{fig:spec_3tm_T-2}.}
\label{fig:equlib_3tm}%
\end{figure}

\begin{figure}
[tbp]
\centering
\includegraphics[height=5.95cm,width=8.0cm]% aspect ratio: 1.333
{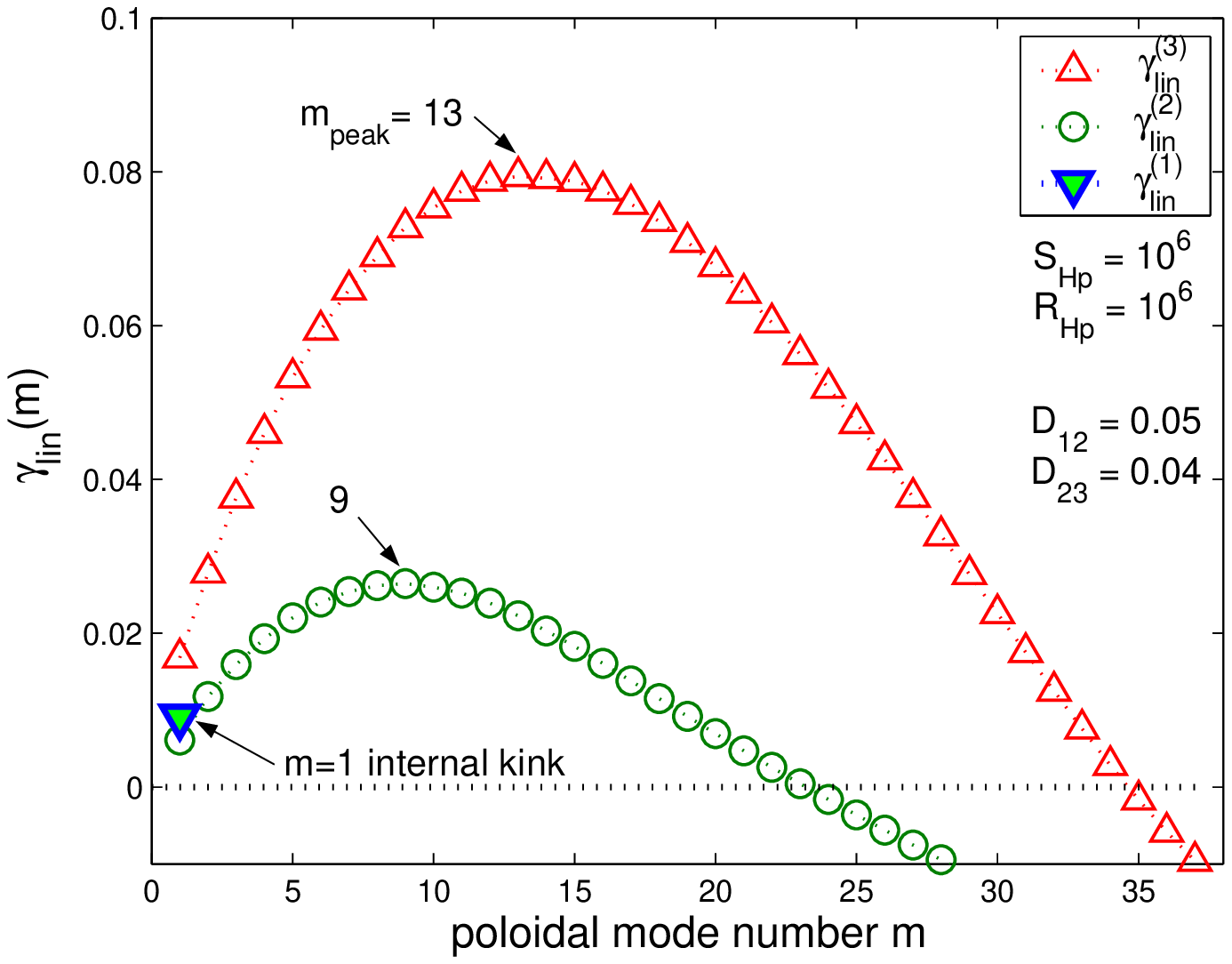}
\caption{(Color online). Spectra $\gLin(m)$ of unstable eigenmodes in Case (T-1) for $\SHp = 10^6$ and $\RHp = 10^6$. The growth rates $\gLin^{(1)}$, $\gLin^{(2)}$ and $\gLin^{(3)}$ of the three eigenmodes $\Ma$, $\Mb$ and $\Mc$ (cf.~Fig.~4 in Ref.~\protect\cite{Bierwage05b}) are shown.}
\label{fig:spec_3tm_T-1}%
\end{figure}

\begin{figure}
[tbp]
\centering
\includegraphics[height=5.95cm,width=8.0cm]% aspect ratio: 1.333
{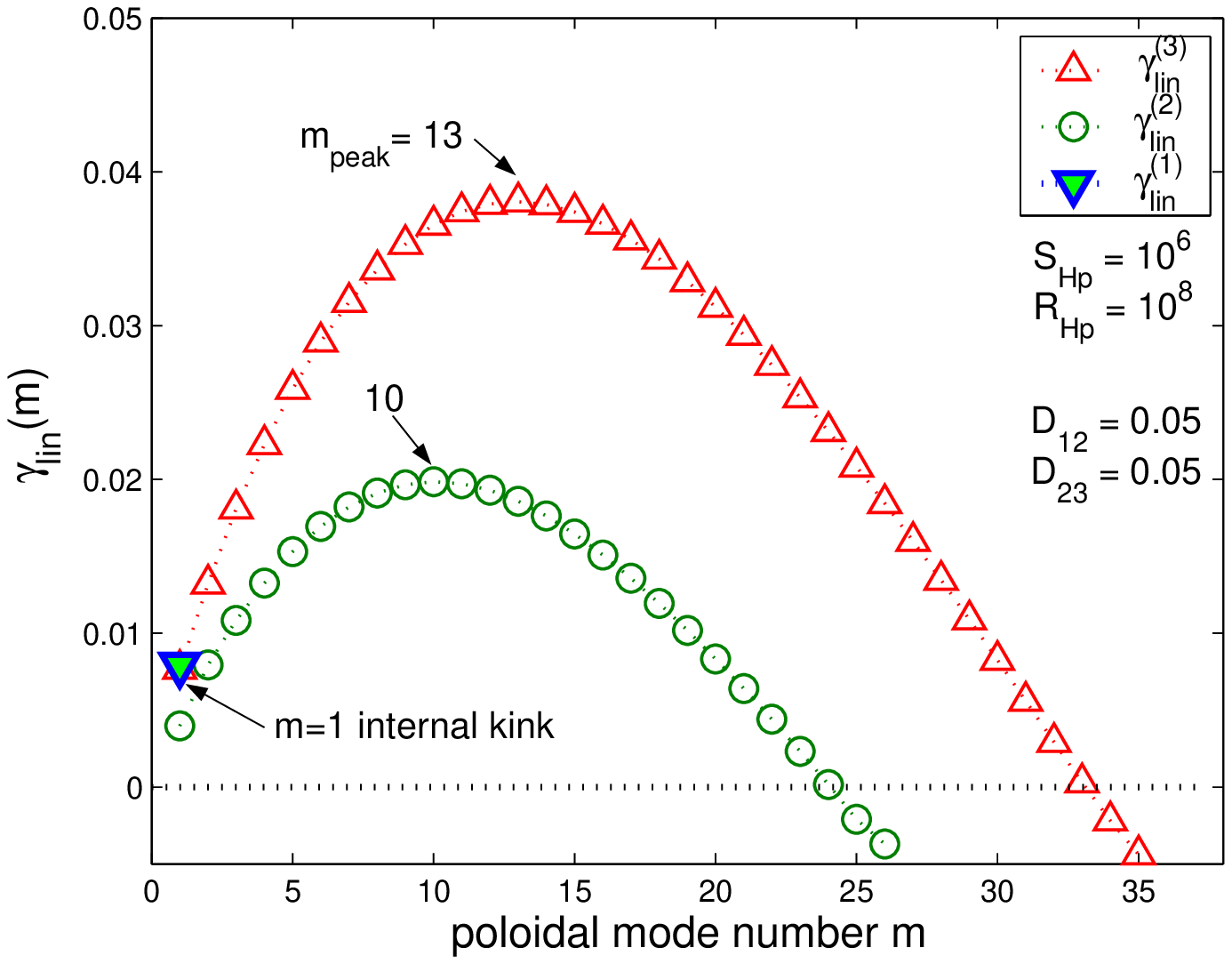}
\caption{(Color online). Spectra $\gLin(m)$ of unstable eigenmodes in Case (T-2) for $\SHp = 10^6$ and $\RHp = 10^8$. The growth rates $\gLin^{(1)}$, $\gLin^{(2)}$ and $\gLin^{(3)}$ of the three eigenmodes $\Ma$, $\Mb$ and $\Mc$ (cf.~Fig.~4 in Ref.~\protect\cite{Bierwage05b}) are shown.}
\label{fig:spec_3tm_T-2}%
\end{figure}

\begin{table}
[tbp]
\centering
\begin{ruledtabular}
\begin{tabular}[t]{c|ccccccccc}
Case & $q_0$ & $r_{\rm A}$ & $\mu_0$ & $\mu_1$ & $m$ & $n$ & $f_1$ & $r_{11}$ & $r_{12}$
\\
\hline (T-1) & $0.73$ & $0.455$ & $0.93$ & $1.45$ & $1$ & $1$ & $-0.09$ & $0.5406$ & $0.039$
\\
\hline (T-2) & $0.73$ & $0.455$ & $0.93$ & $1.45$ & $1$ & $1$ & $-0.098$ & $0.5666$ & $0.064$
\end{tabular}
\end{ruledtabular}
\caption{Parameter values for the $q$ profiles shown in Fig.~\protect\ref{fig:equlib_3tm}, using model formula (11) in Ref.~\protect\cite{Bierwage05b}.}
\label{tab:equlib_3tm_q-parm}
\end{table}

\begin{table}
[tbp]
\centering
\begin{ruledtabular}
\begin{tabular}{c|ccccccccc}
Case & $D_{12}$ & $D_{23}$ & $s_1$ & $s_2$ & $s_3$ & $\eta_1$ & $\eta_2$ & $\eta_3$ & $\mPeak$
\\
\hline (T-1) & $0.053$ & $0.043$ & $0.35$ & $-0.56$ & $1.20$ & $1.7$ & $1.1$ & $3.5$ & $13$
\\
\hline (T-2) & $0.052$ & $0.052$ & $0.23$ & $-0.23$ & $0.59$ & $1.6$ & $1.2$ & $1.9$ & $13$
\end{tabular}
\end{ruledtabular}
\caption{Properties of the $q$ profiles shown in Fig.~\protect\ref{fig:equlib_3tm}. The mode number of the fastest growing mode, $\mPeak$ (cf. Figs.~\protect\ref{fig:spec_3tm_T-1} and \protect\ref{fig:spec_3tm_T-2}), is valid for $\SHp = 10^6$ and $\RHp = 10^6$ [Case (T-1)], $\RHp = 10^8$ [Case (T-2)].}
\label{tab:equlib_3tm}
\end{table}

\begin{figure}
[tbp]
\centering
\includegraphics[height=11.956cm,width=8.0cm]% aspect ratio: 0.669
{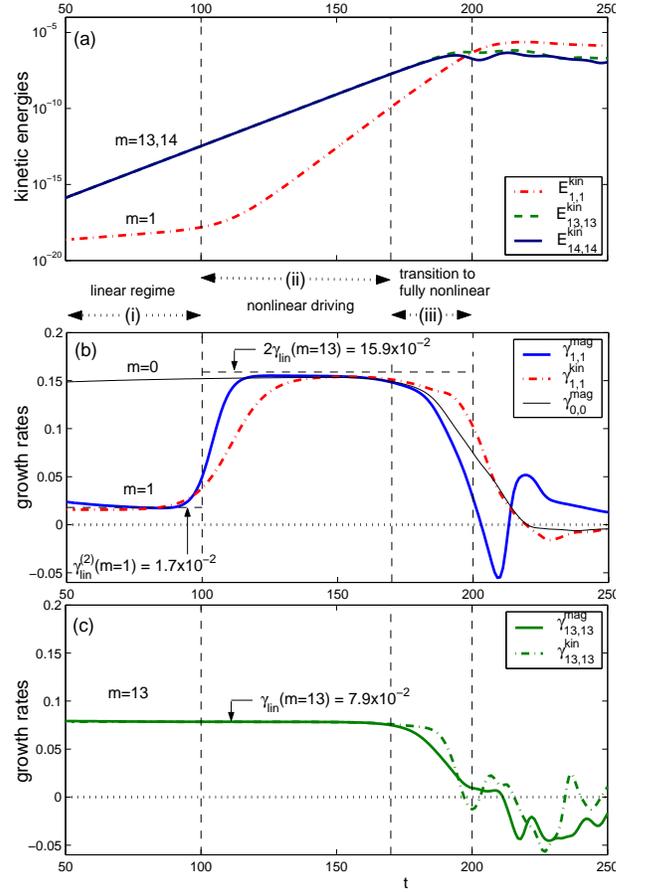}
\caption{(Color online). Early evolution in Case (T-1). (a): Kinetic energies $E^{\rm kin}_{m,n}$ [Eq.~(\protect\ref{eq:def-E})] of the $m=1$ mode and the two fastest growing modes $m=13$ and $m=14$. (b) and (c): Magnetic and kinetic growth rates $\gamma^{\rm mag}_{m,n}$ and $\gamma^{\rm kin}_{m,n}$ [Eq.~(\protect\ref{eq:def-gr})]. The main stages to be distinguished are: (i) linear growth, (ii) nonlinearly driven growth, (iii) transition to the fully nonlinear (turbulent) regime. The value of the driven growth rate expected during stage (ii) is $\gDrive \approx 2\gLin(\mPeak) \approx 16\times 10^{-3}$, as indicated in (b). $\SHp = 10^6$, $\RHp = 10^6$.}
\label{fig:E-g_3tm1_T-1}%
\end{figure}

\begin{figure}
[tbp]
\centering
\includegraphics[height=5.984cm,width=8.0cm]% aspect ratio: 1.337
{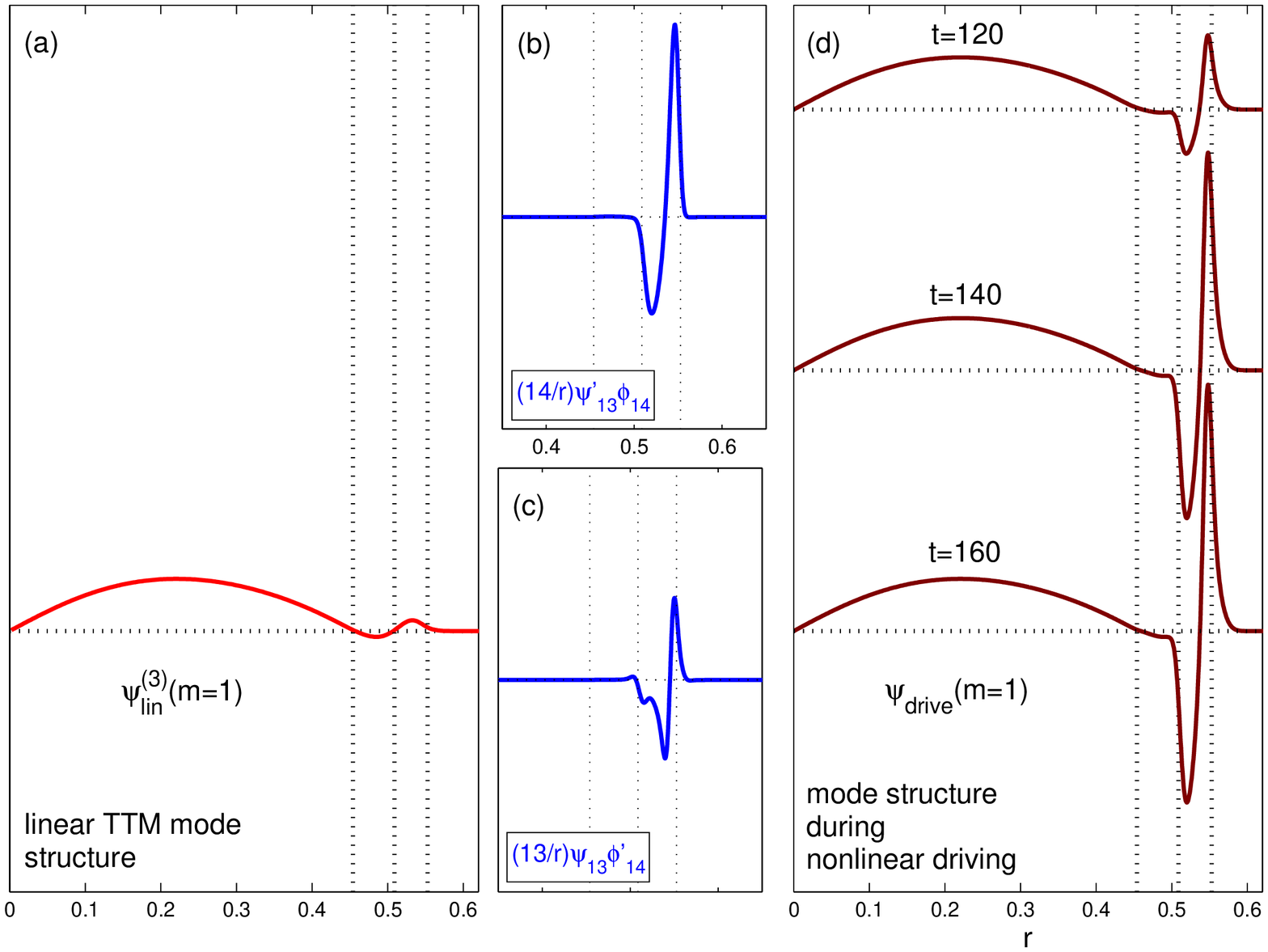}
\caption{(Color online). Effect of the nonlinear driving on the $m=1$ mode structure. In (a) the linear mode structure of the fastest growing $m=1$ eigenmode, $\psi_{\rm lin}^{(3)}(m=1)$, is shown. The two fastest growing modes, $\mPeak=13$ and $\mPeak+1=14$ (cf.~Fig.~\protect\ref{fig:spec_3tm_T-1}), give rise to a rapidly growing radially localized $m=1$ component through the convective nonlinearity $[\psi,\phi]$ in Eq.~(\protect\ref{eq:rmhd1}). In (b) and (d) the structures of the two main terms are plotted [(b) dominates]. In (d) we show the situation before ($t=120$) and after the new fast growing $m=1$ mode structure is established ($t=140$ and $180$). The curves in (a) and (d) are scaled such that the amplitude of the internal kink component ($0 < r \leq \rsa$) is the same at all times. Resonant surfaces are indicated by dotted vertical lines.}
\label{fig:nl_drive}%
\end{figure}

The equilibrium state is taken to be axisymmetric (only $m=n=0$ components) and free of flows, i.e.,
\begin{equation}
\eqPhi = \eqVor = 0.
\end{equation}

\noindent The equilibrium magnetic configuration is uniquely defined in terms of the safety factor $q(r)$, and the magnetic flux function and current density profiles are obtained though the relations
\begin{equation}
q^{-1} = -\frac{1}{r} \frac{{\rm d}}{{\rm d}r}\psi_{0,0} \quad {\rm and} \quad j_{0,0} = \frac{1}{r} \frac{{\rm d}}{{\rm d}r} \frac{r^2}{q}.
\label{eq:q-equlib}
\end{equation}

In this paper configurations with three resonant surfaces with $\qRes \equiv q(\rsi) = m/n = 1$, which are located at radii $\rsi$ ($i=1,2,3$), are considered. The distances between the resonances, $D_{ij} = |\rsj - \rsi|$, are chosen sufficiently small, so that the spectra of unstable TTMs are broad and the dominant modes have $m \sim \mathcal{O}(10)$. The equilibrium $q(r)$ profiles used are shown in Fig.~\ref{fig:equlib_3tm}. They are obtained using the model formula given by Eq.~(11) in Ref.~\cite{Bierwage05b} with the parameters in Table~\ref{tab:equlib_3tm_q-parm}. The two cases studied are labeled (T-1) and (T-2).

The dispersion relations (spectra of linear growth rates) $\gLin(m)$ for all unstable eigenmodes are given in Fig.~\ref{fig:spec_3tm_T-1} for Case (T-1) and in Fig.~\ref{fig:spec_3tm_T-2} for Case (T-2). The linear eigenmode structures for Case (T-1) were shown in Ref.~\protect\cite{Bierwage05a} and are similar for Case (T-2). Let us recall that eigenmode $\Ma$ (with growth rate $\gLin^{(1)}$) is associated with the resonant surface $r=\rsa$ and is an ordinary (single) $m=1$ internal kink-tearing mode (stable for $m>1$). $\Mb$ (with $\gLin^{(2)}$) is associated with $\rsb$ in the sense that it is active in the region $0 < r < \rsb$ for $m=1$ and  $\rsa < r < \rsb$ for $m>1$. Thus, it may be regarded as a double tearing mode (DTM). Finally, $\Mc$ (with $\gLin^{(3)}$) is associated with $\rsc$ and may be regarded as the actual TTM eigenmode. In Figs.~\ref{fig:spec_3tm_T-1} and \ref{fig:spec_3tm_T-2} the dominant eigenmodes and the $m=1$ mode are indicated by arrows labeled with the corresponding mode numbers.

\begin{figure*}
[tbp]
\centering
\includegraphics[height=9.787cm,width=17cm]% aspect ratio: 1.737
{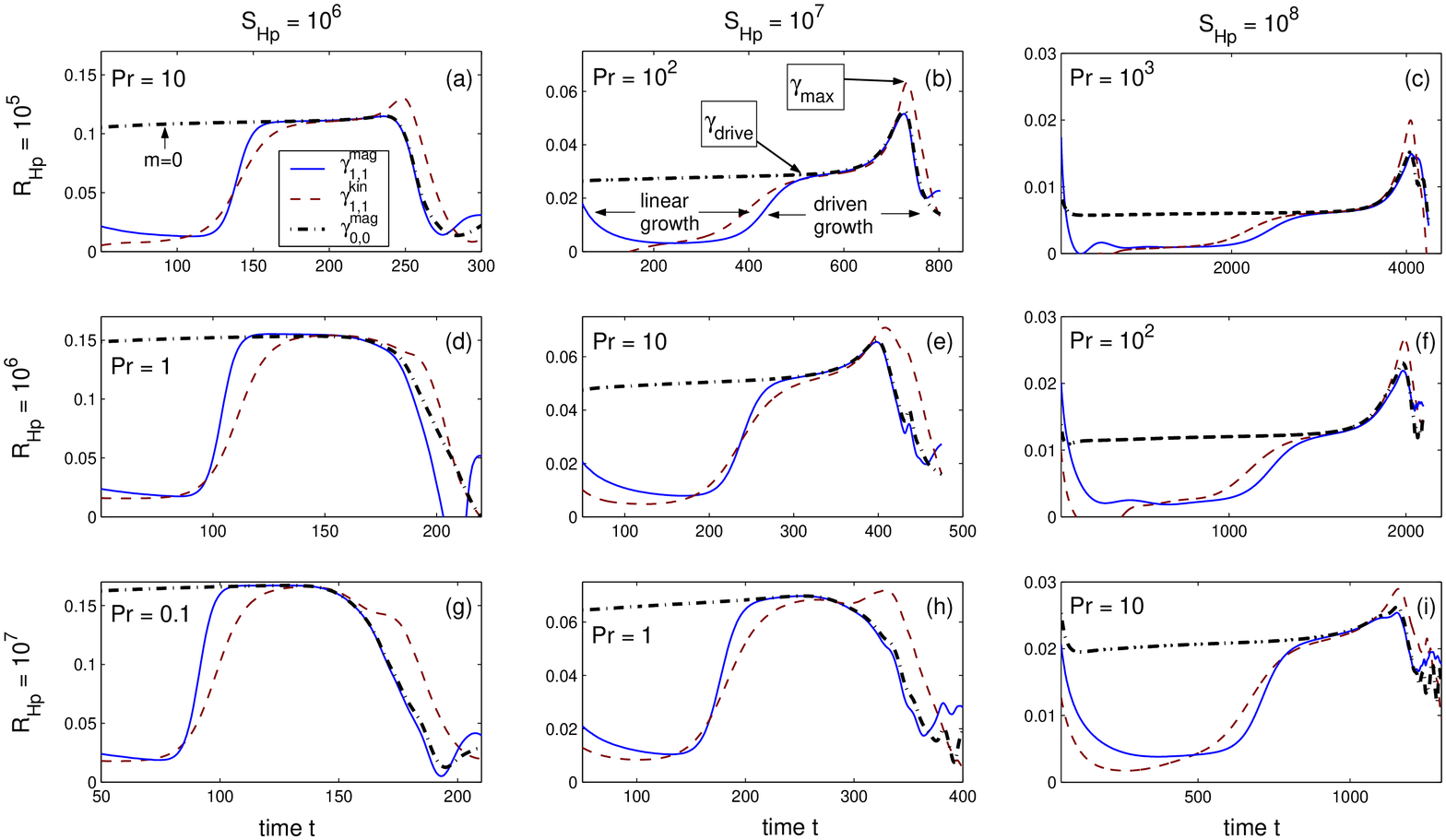}
\caption{(Color online). Effect of varying $\SHp$ and $\RHp$ on the evolution of the $m=1$ mode, described in terms of $\gKin_{1,1}$ (dashed line) and $\gMag_{1,1}$ (solid line). The growth rate of the $m=0$ mode is plotted as well (dash-dotted line), showing the level of nonlinear driving starting at $t=0$. We distinguish between linear growth ($\gLin$) and nonlinearly driven growth due to the fastest growing modes ($\gDrive$). For $Pr \geq 10$ explosive growth follows, up to a maximum growth rate $\gMax$, as indicated in (b). The time interval shown is from $t=50$ (after the transient relaxation) until the annular collapse and onset of MHD turbulence. The configuration used is Case (T-1). The data for $Pr = 1$ in (d) is the same as in Fig.~\protect\ref{fig:E-g_3tm1_T-1}(b) above and in Fig.~5 of Ref.~\protect\cite{Bierwage05a}.}
\label{fig:comp_E-g_S-R}%
\end{figure*}

The characteristics of the equilibrium configurations are summarized in Table~\ref{tab:equlib_3tm}, including the values of the resistivity profile $\hat{\eta}$ at the resonances and the poloidal mode number $\mPeak$ of the dominant TTM eigenmode obtained from Figs.~\ref{fig:spec_3tm_T-1} and \ref{fig:spec_3tm_T-2}. We will see that the value of $\mPeak$ is an important clue for the interpretation of the nonlinear dynamics since it dictates a structure size that is most likely to be encountered in the poloidal direction.

Starting from an unstable equilibrium, the instability is excited by applying an initial perturbation of the form
\begin{equation}
\flPsi(t=0) = \frac{1}{2}\sum\limits_{m=1}^{31}\Psi_{0,m} r (r-1) e^{im(\vartheta_* + \vartheta_{0,m})} + {\rm c.c},
\label{eq:pert}
\end{equation}

\noindent where $\Psi_{0,m} = 10^{-11}$ is the perturbation amplitude and $\vartheta_* \equiv \vartheta - \qRes^{-1}\zeta$ is a helical angle coordinate. Each mode has an initial poloidal phase shift $\vartheta_{0,m}$ with a randomly assigned value in the range $\vartheta_{0,m} \in [0, \pi]$.

%===========================================
\section{Early dynamics: establishment of nonlinear mode structure and fast growth of the $m=1$ mode}
\label{sec:kink}

Due to the disparate growth rates between the $m=1$ mode and the fastest growing TTMs [here, $\gPeak \sim 5\times\gLin(m=1)$; cf. Figs.~\ref{fig:spec_3tm_T-1} and \ref{fig:spec_3tm_T-2}] nonlinear interactions begin already in regimes where mode amplitudes are still small. In this section we focus on these early stages of evolution where turbulence has not yet developed. They are most conveniently studied by considering the dynamics of individual Fourier modes. Results are presented only for Case (T-1), using $\SHp = 10^6$ and $\RHp = 10^6$. Similar behavior is observed in Case (T-2).

Figure~\ref{fig:E-g_3tm1_T-1}(a) shows the evolution of the kinetic energies of the $m=1$ mode and the two fastest growing modes $\mPeak=13$ and $m=14$ (cf.~Fig.~\ref{fig:spec_3tm_T-1}). The corresponding growth rates are shown in Fig.~\ref{fig:E-g_3tm1_T-1}(b) and (c). It can be seen that the $m=13$ and $m=14$ modes grow at their linear growth rates until $t\approx 170$ and saturate during $170 \lesssim t \lesssim 200$. The $m=1$ mode grows linearly until $t\approx 100$ [phase (i) in Fig.~\ref{fig:E-g_3tm1_T-1}]. Subsequently, its growth rate increases and exponential growth continues at an enhanced rate during $100 \lesssim t \lesssim 170$ [phase (ii)]. Upon entering the nonlinear regime, the $m=1$ growth rate drops [phase (iii)].

\begin{figure*}
[tbp]
\centering
\includegraphics[height=18.0cm]% aspect ratio: 0.814
{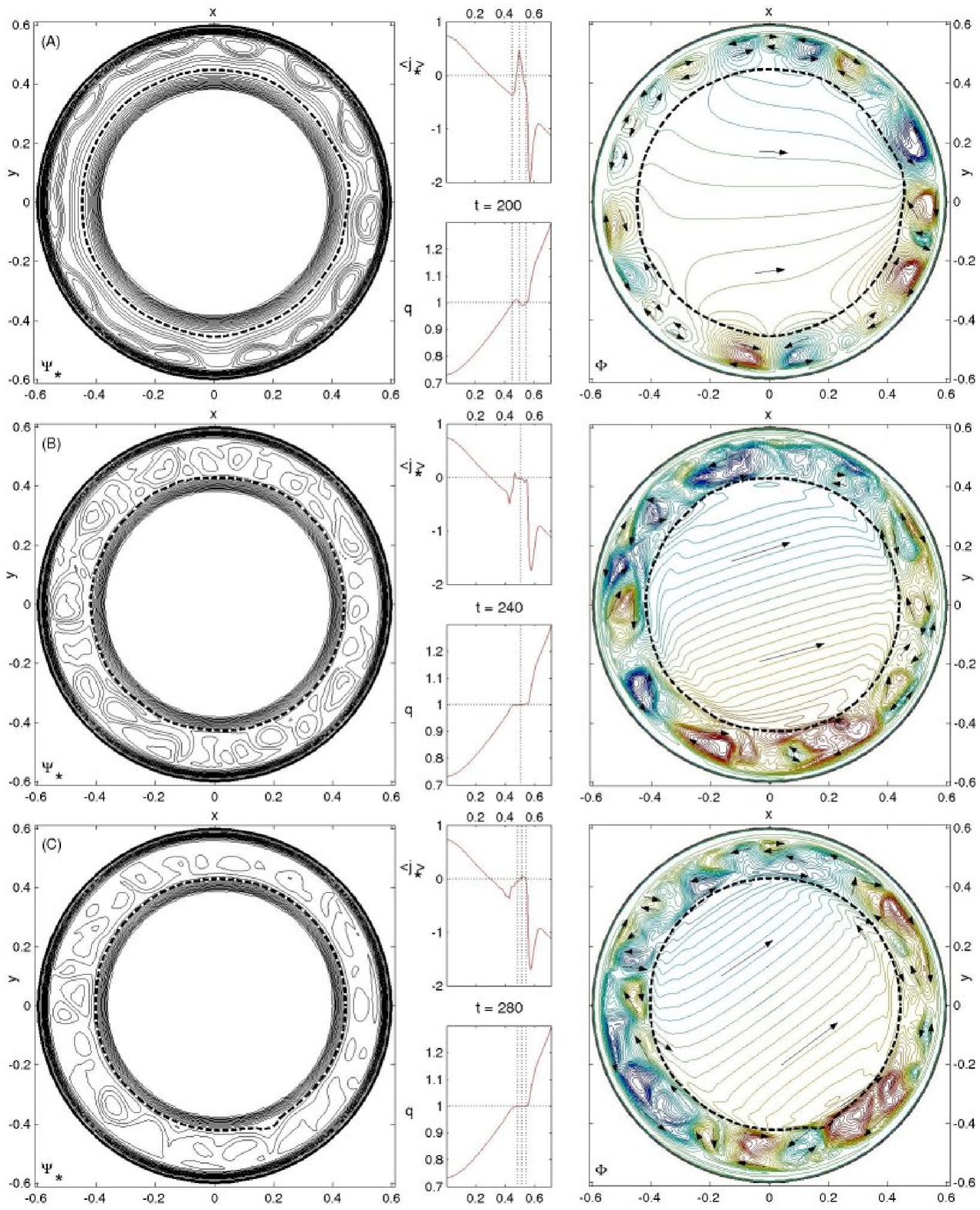}
\caption{(Color online). Annular collapse and long-term evolution in Case (T-1). The first three snapshots taken at (A) $t=200$, (B) $t=240$, and (C) $t=280$ are shown (continued in Fig.~\protect\ref{fig:snaps_3tm_T-1b}). Each snapshot consists of contour plots of the helical flux $\psi_*$ (left) and the electrostatic potential $\phi$ (right). Arrows indicate the flow directions. The dashed circles indicate the outermost $\psi$ contour of the core and have been superimposed on the $\phi$ contours for clarity. The small diagrams in the middle show the instantaneous profiles $q(r,t)$ and $\left<j_*\right> \equiv [j_*(r,t)]_{0,0}$. $\SHp = 10^6$, $\RHp = 10^6$.}
\label{fig:snaps_3tm_T-1a}%
\end{figure*}

\begin{figure*}
[tbp]
\centering
\includegraphics[height=18.0cm]% aspect ratio: 0.814
{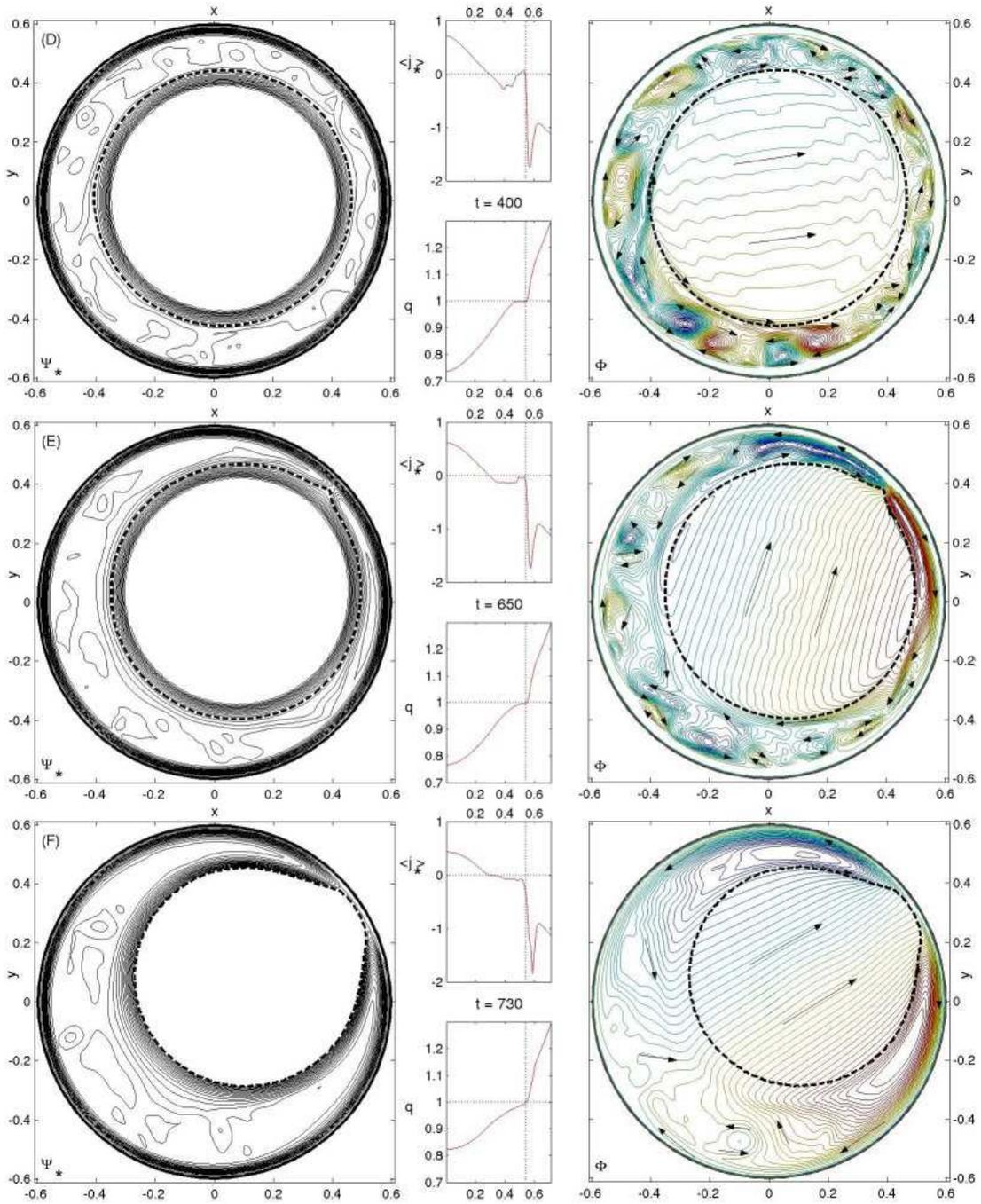}
\caption{(Color online). Long-term evolution in Case (T-1). Continuing Fig.~\protect\ref{fig:snaps_3tm_T-1a}, snapshots were taken at (D) $t=400$, (E) $t=650$, and (F) $t=730$. Arranged as Fig.~\protect\ref{fig:snaps_3tm_T-1a}. $\SHp = 10^6$, $\RHp = 10^6$.}
\label{fig:snaps_3tm_T-1b}%
\end{figure*}

The enhanced growth during phase (ii) is due to nonlinear driving by the fastest growing modes, predominantly $\mPeak=13$ and $\mPeak\pm 1$. Since $\gLin(\mPeak) \approx \gLin(\mPeak\pm 1)$ the driven growth rate of the $m=1$ mode is $\gDrive \approx 2\gLin(\mPeak) \approx 16\times 10^{-2}$. The purely nonlinearly driven $m=0$ mode grows at the same rate, as can be seen in Fig.~\ref{fig:E-g_3tm1_T-1}(b) [phases (i) and (ii)]. Note that in the present case $\gDrive$ this is almost one order of magnitude higher than the linear growth rate of the $m=1$ mode, $\gLin(m=1) = 1.7\times 10^{-2}$.

The effect of the nonlinear driving on the $m=1$ mode structure can be seen in Fig.~\ref{fig:nl_drive}. In Fig.~\ref{fig:nl_drive}(a) the linear mode structure of the TTM-type eigenmode is plotted, specifically the flux function $\psi_{\rm lin}^{(3)}(m=1)$. The shapes of the driving terms arising from the convective nonlinearity $[\psi,\phi]$ in Eq.~(\ref{eq:rmhd1}) are plotted in Fig.~\ref{fig:nl_drive}(b) and (c). In Fig.~\ref{fig:nl_drive}(d) it can be seen how this radially localized driving appears in the nonlinear $m=1$ mode structure $\psi_{\rm drive}(m=1)$ in the form of ``spikes'' located near the resonant surfaces. It is particularly remarkable that, after a certain ratio between the amplitudes of the driving component (around $\rsc$) and the component remaining from the linear eigenmode structure (in the region $0 < r < \rsa$) is reached, the mode structure does not change further [Fig.~\ref{fig:nl_drive}(d), $t=140$--$160$] and grows \emph{as a whole} at the enhanced rate $\gDrive$. Thus, a new fast growing global $m=1$ mode is created through radially localized driving. Comparisons made with other simulation runs indicate that the relative size of the driving component and the peak in the region $0 < r < \rsa$ in the established nonlinear mode structure depends on the ratio between $\gDrive$ and $\gLin(m=1)$: If $\gDrive/\gLin(m=1)$ is large, the driving component (around $\rsc$) has a large amplitude compared to the kink component ($0 < r \lesssim \rsa$), and vice versa.

%===========================================
\section{Behavior of the $m=1$ mode during the transition to turbulence}
\label{sec:trans}

Let us now consider the transition to the fully nonlinear, turbulent regime. In Fig.~\ref{fig:E-g_3tm1_T-1} this transition takes place via a gradual decrease in the growth rates during phase (iii). However, it turns out that this is only the case in regimes where the effect of viscosity is negligible. In Fig.~\ref{fig:comp_E-g_S-R} the evolution of the nonlinear growth rates $\gMag_{1,1}$, $\gKin_{1,1}$ and $\gMag_{0,0}$ is shown for magnetic Reynolds numbers $\SHp = 10^6, 10^7, 10^8$ (from left to right) and the kinematic Reynolds numbers $\RHp = 10^5, 10^6, 10^7$ (from top to bottom). Thus, the Prandtl number $Pr$, which measures the relative strengths of viscosity and (resistive) diffusion ($Pr = \SHp/\RHp \propto \nu/\eta_0$) varies over the range $10^{-1} \leq Pr \leq 10^3$ (bottom left to top right). The growth rate of the $m=0$ mode, $\gMag_{0,0}$, is shown since it clearly reflects the level of nonlinear driving at all times. Qualitatively, the results in Fig.~\ref{fig:comp_E-g_S-R} may be summarized as follows:

\begin{enumerate}
\item  $Pr \sim 0.1$ [Fig.~\ref{fig:comp_E-g_S-R}(g)]: During the transition to the nonlinear regime, $\gKin_{1,1}$ first decreases slowly ($t\approx 150$--$180$) and then drops rapidly. Similar behavior is observed for $Pr = 10^{-2}$ and is also expected for $Pr < 10^{-2}$.

\item  $Pr \sim 1$ [Fig.~\ref{fig:comp_E-g_S-R}(d,h)]: Compared to the results obtained with lower $Pr$, $\gDrive$ is now reduced due to the stabilizing effect of the viscosity on $\gLin(m)$ (cf. Fig.~5 in Ref.~\cite{Bierwage05b}). During the transition to the fully nonlinear regime $\gKin_{1,1}$ remains high for a certain period of time and may exhibit oscillatory behavior as in Fig.~\ref{fig:comp_E-g_S-R}(h) before dropping.

\item  $Pr \sim 10, 10^2, 10^3$ [Fig.~\ref{fig:comp_E-g_S-R}(a,b,c,e,f,i)]: The growth rate $\gDrive$ is further decreased due to higher viscosity. However, at the end of the driving phase a significant increase in the growth rate from $\gDrive \approx 2\gPeak$ to a value $\gMax$ [indicated in Fig.~\ref{fig:comp_E-g_S-R}(b)] is observed. This effect is most pronounced in the kinetic growth rate $\gKin_{1,1}$. Note in particular that $\gMax$ can get close to the value of $\gDrive$ obtained for $Pr \sim 1$ [compare, e.g., Fig.~\ref{fig:comp_E-g_S-R}(b), (e) and (h)]. It is likely that this behavior can also be observed for $Pr > 10^3$.
\end{enumerate}

Let us note that during this rapid growth the nonlinear interactions already include many Fourier modes. We suspect that the explosive growth phase observed here for $Pr \gtrsim 10$ is due to these nonlinear interactions becoming more important than viscous damping. The latter had reduced $\gDrive$ through a reduction of the linear growth rates $\gLin(m)$.

%===========================================
\section{Annular collapse and effect of MHD turbulence on the internal kink}
\label{sec:recon}

In this section we investigate the long-term evolution in Cases (T-1) and (T-2) (Fig.~\ref{fig:equlib_3tm}). We describe the magnetic and $\ExB$ flow structures generated through TTM reconnection and analyze the evolution of the $m=1$ internal kink mode while it is surrounded by MHD turbulence. The values for the dissipation parameters in Case (T-1) are $\SHp = 10^6$ and $\RHp = 10^6$, as in Sec.~\ref{sec:kink}. For Case (T-2) we choose $\SHp = 10^6$ and $\RHp = 10^8$.

\begin{figure}
[tb]
\centering
\includegraphics[height=11.956cm,width=8.0cm]% aspect ratio: 0.669
{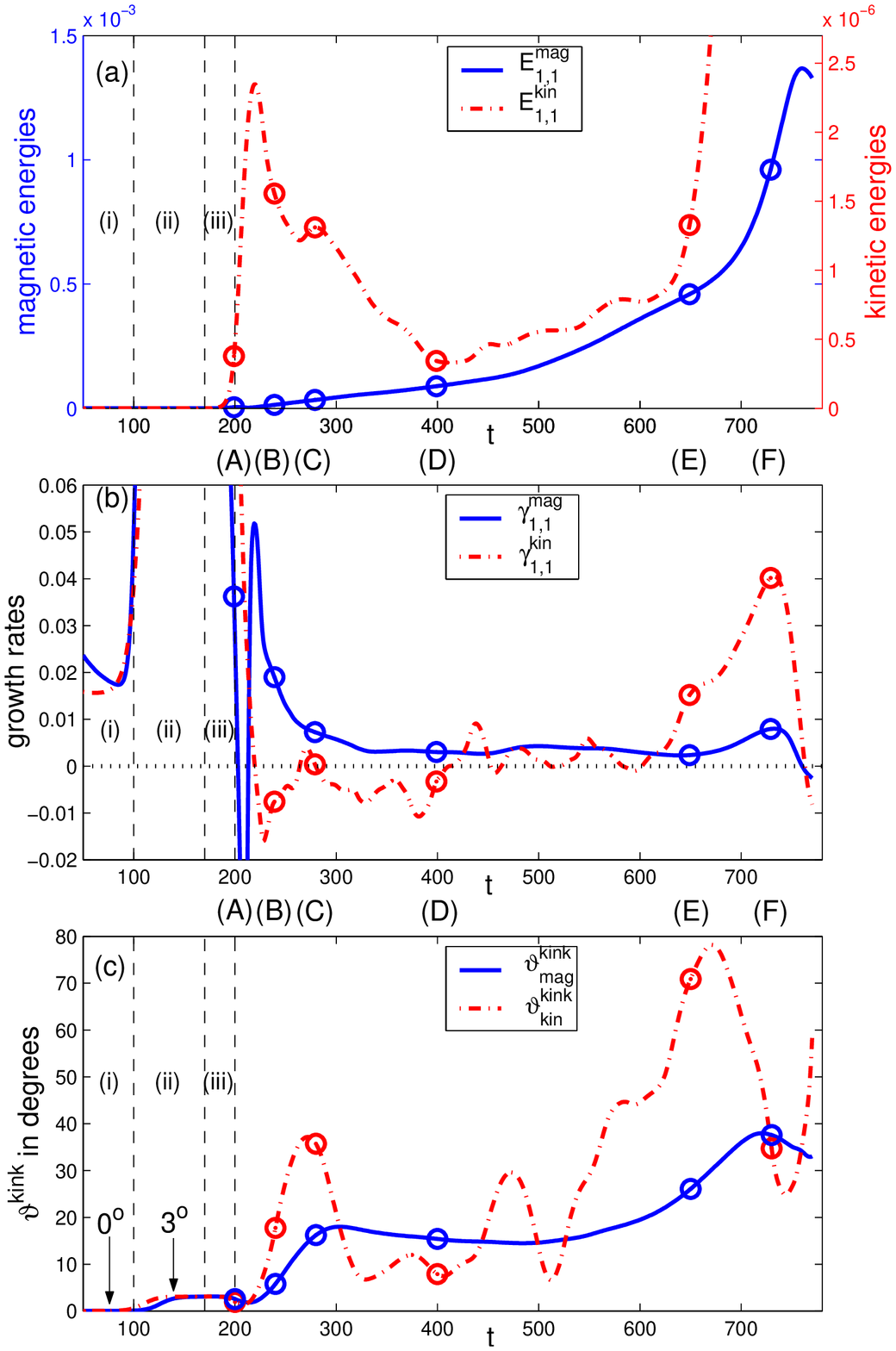}
\caption{(Color online). Long-term evolution of the $m=1$ mode in Case (T-1). (a): Magnetic and kinetic energies $\EMag_{1,1}$ and $\EKin_{1,1}$. (b): Magnetic and kinetic growth rates $\gMag_{1,1}$ and $\gKin_{1,1}$. (c): Motion of the core in the kink flow with dominant mode number $m=1$. The magnetic axis is located at the poloidal angle $\vartheta^{\rm kink}_{\rm mag}$ and is moving in the direction $\vartheta^{\rm kink}_{\rm kin}$ (in the poloidal plane at $\zeta = 0$) [see definitions in Eq.~(\protect\ref{eq:ph-kink})]. In (a), (b) and (c) the times at which snapshots were taken are indicated by circles and labeled (A)--(F) [cf.~Figs.~\protect\ref{fig:snaps_3tm_T-1a} and \protect\ref{fig:snaps_3tm_T-1b}].}
\label{fig:E-g_T-1_long}%
\end{figure}

\begin{figure*}
[tbp]
\centering
\includegraphics[height=18.0cm]% aspect ratio: 0.814
{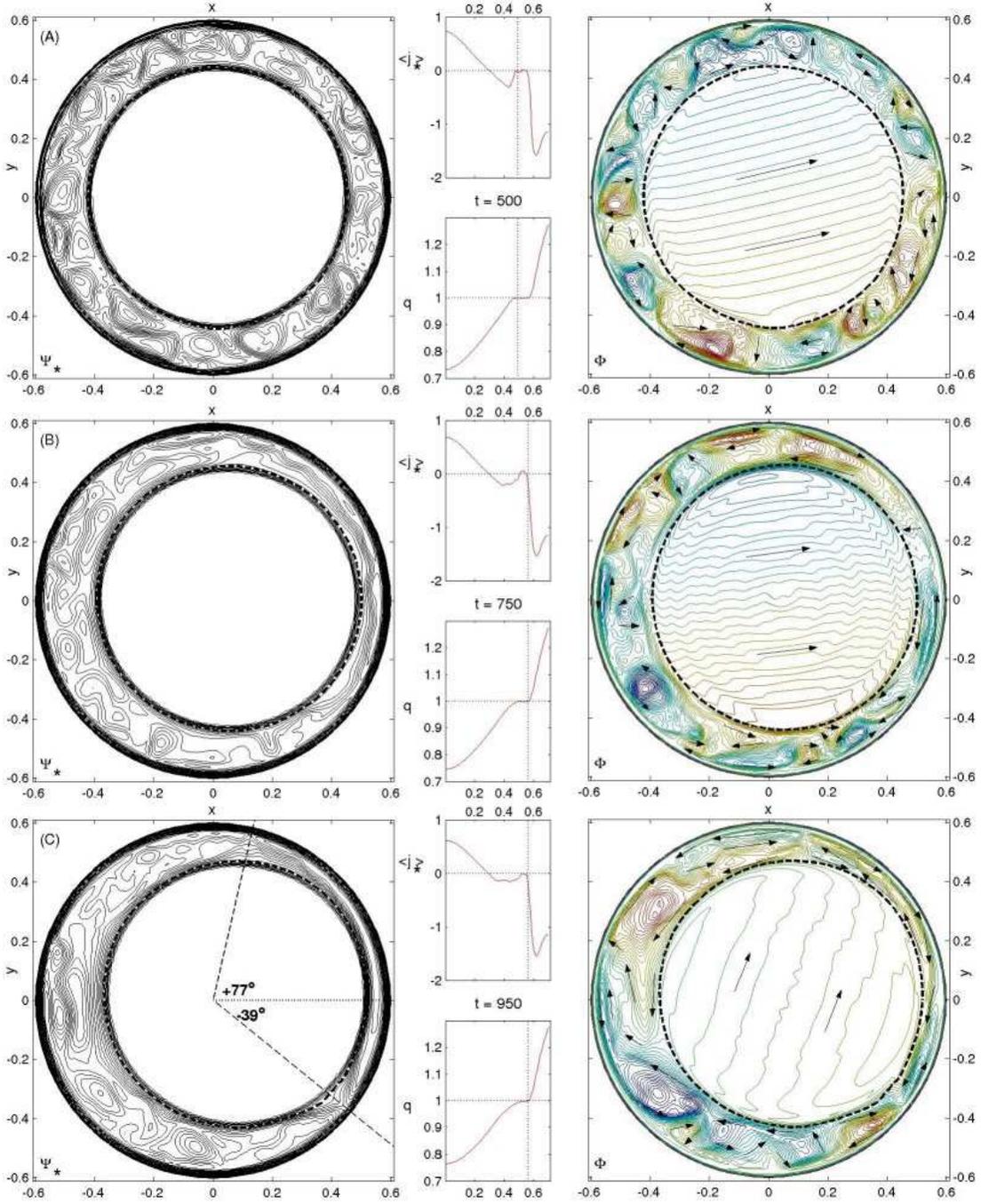}
\caption{(Color online). Annular collapse and long-term evolution in Case (T-2). The first three snapshots taken at (A) $t=500$, (B) $t=750$, and (C) $t=950$ are shown (continued in Fig.~\protect\ref{fig:snaps_3tm_T-2b}). $\SHp = 10^6$, $\RHp = 10^8$.}
\label{fig:snaps_3tm_T-2a}%
\end{figure*}

\begin{figure*}
[tbp]
\centering
\includegraphics[height=18.0cm]% aspect ratio: 0.814
{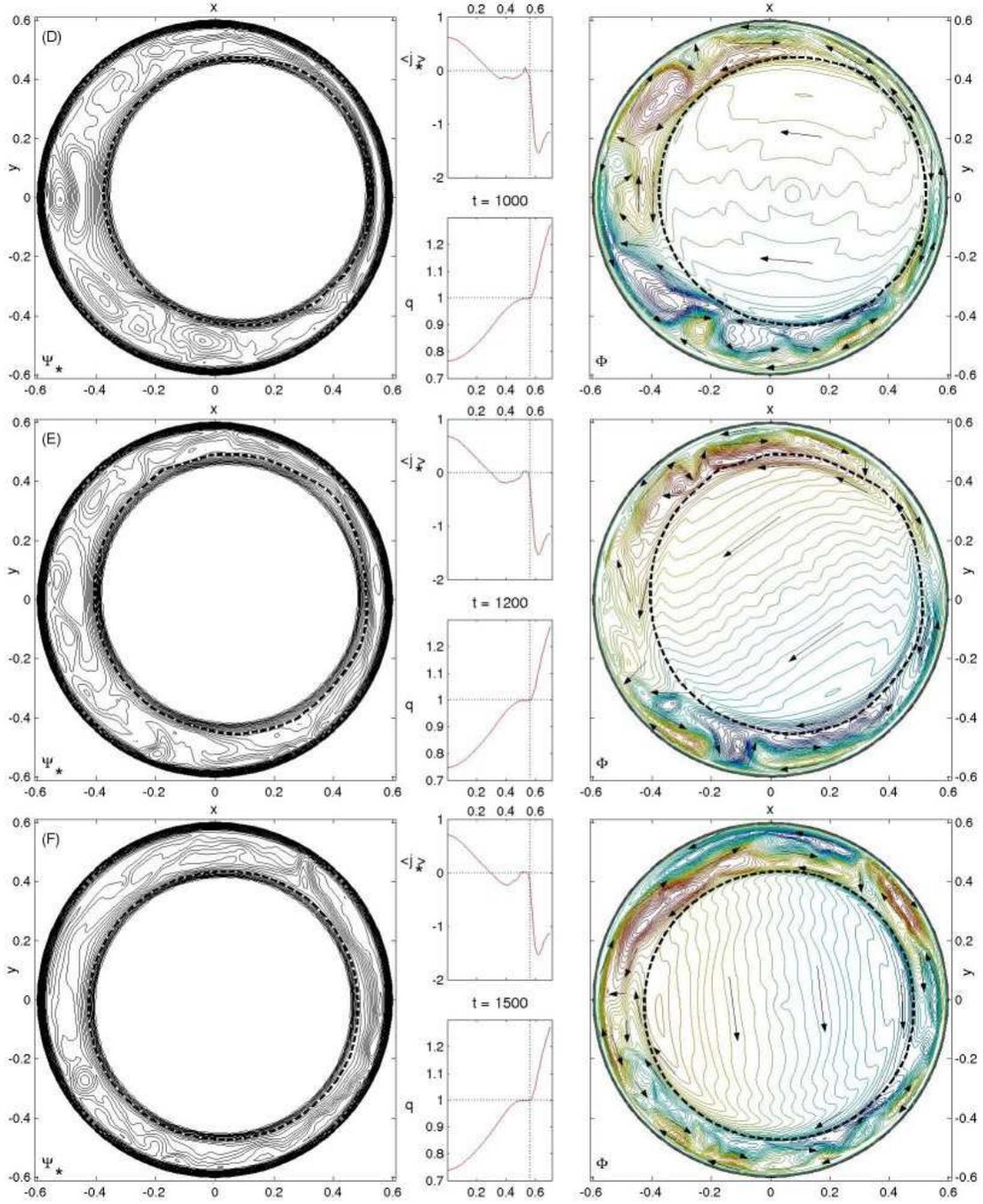}
\caption{(Color online). Long-term evolution in Case (T-2). Continuing Fig.~\protect\ref{fig:snaps_3tm_T-2a}, snapshots were taken at (D) $t=1000$, (E) $t=1200$, and (F) $t=1500$. Arranged as Fig.~\protect\ref{fig:snaps_3tm_T-1a}. $\SHp = 10^6$, $\RHp = 10^8$.}
\label{fig:snaps_3tm_T-2b}%
\end{figure*}

\begin{figure}
[tb]
\centering
\includegraphics[height=11.956cm,width=8.0cm]% aspect ratio: 0.669
{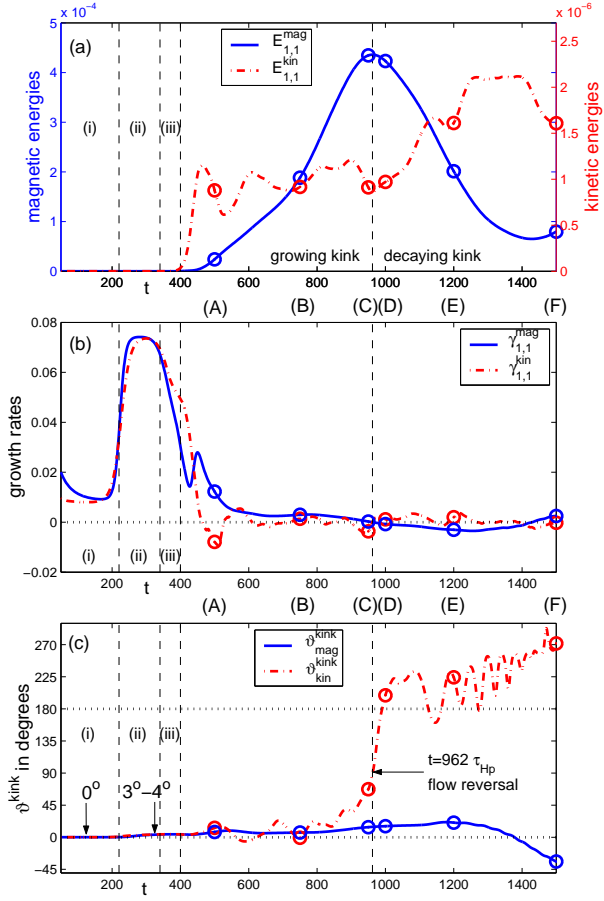}
\caption{(Color online). Long-term evolution of the $m=1$ mode in Case (T-2). Arranged as Fig.~\protect\ref{fig:E-g_T-1_long}. In (a), (b) and (c) the times at which snapshots were taken are indicated by circles and labeled (A)--(F) [cf.~Figs.~\protect\ref{fig:snaps_3tm_T-2a} and \protect\ref{fig:snaps_3tm_T-2b}].}
\label{fig:E-g_T-2_long}%
\end{figure}

\begin{figure}
[tb]
\centering
\includegraphics[height=11.956cm,width=8.0cm]% aspect ratio: 0.669
{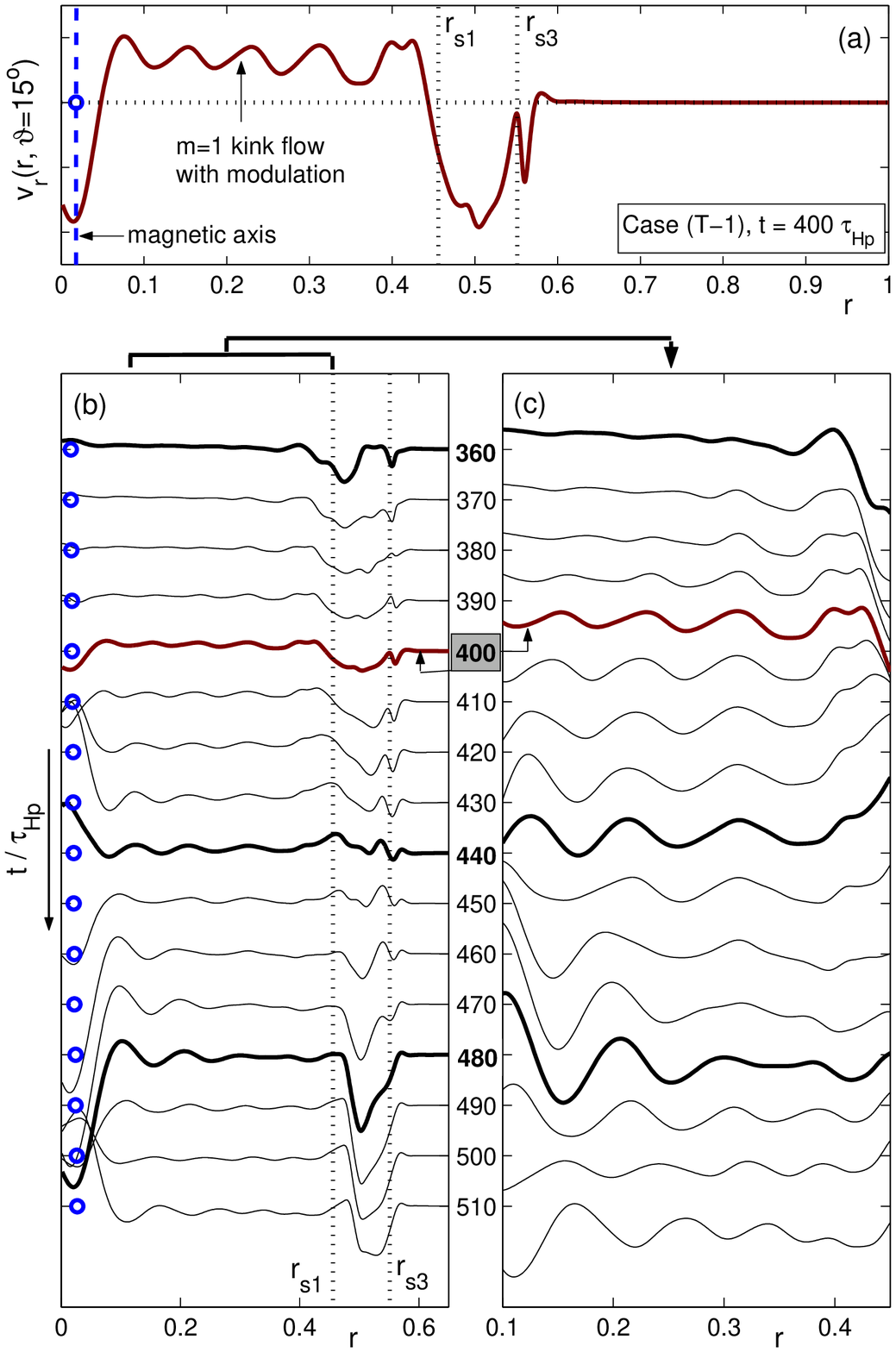}
\caption{(Color online). Evolution of the $m=1$ modulation on the radial displacement velocity profile $v_r \propto \phi/r$ in Case (T-1). (a): Complete $v_r$ profile along the poloidal angle $\vartheta = 15^o$, i.e., roughly where the magnetic axis is located. The time is $t=400$ and coincides with Snapshot~(D) in Fig.~\protect\ref{fig:snaps_3tm_T-1b}. (b): Temporal evolution of the $m=1$ modulation in the interval $360 \leq t \leq 510$ (from top to bottom). The radial location of the magnetic axis varies as indicated by the circles [e.g., $r_{\rm axis}(t=360)=1.6\times 10^{-2}$, and $r_{\rm axis}(t=510) = 2.7\times 10^{-2}$]. The average radial velocity of the core is about $\overline{v}_r \sim 0.7\times 10^{-2}$ (in $a/\tHp$). (c): Same data as in (b) but in the radial interval $0.1 \leq r \leq 0.5$ and scaled up for clarity.}
\label{fig:vr_T-1}%
\end{figure}

%-------------------------------------------
\subsection{Case (T-1): Annular collapse and full reconnection}

We begin with a discussion of snapshots taken in Case (T-1), which are shown in Figs.~\ref{fig:snaps_3tm_T-1a} and \ref{fig:snaps_3tm_T-1b} and labeled (A)--(F). The initial perturbation is sufficiently random so that reconnection occurs all around the core [Fig.~\ref{fig:snaps_3tm_T-1a}(A)]. However, it is not isotropic, so that some magnetic islands grow faster than others. The sizes of the islands reflect the shape of the spectrum of linear growth rates $\gLin(m)$ with dominant modes having $m \sim \mPeak = 13$ [cf.~Fig.~\ref{fig:spec_3tm_T-1}]. During this \emph{annular collapse} the amplitude of the $m=1$ mode is still small and the $q(r)$ profile is flattened annularly [Fig.~\ref{fig:snaps_3tm_T-1a}(A)]. The magnetic islands in the inter-resonance region exhibit complicated coalescence dynamics and a \emph{turbulent} annular region is created around the core [Fig.~\ref{fig:snaps_3tm_T-1a}(B) and (C)]. In the meantime, the core displacement becomes observable. The core traverses the turbulent belt [Fig.~\ref{fig:snaps_3tm_T-1b}(D) and (E)] and upon making contact with the flux surfaces beyond the outermost resonant radius ($r = \rsc$) core reconnection begins [Fig.~\ref{fig:snaps_3tm_T-1b}(E) and (F)]. In the present case, (T-1), core reconnection is likely to proceed to completion, i.e., \emph{full reconnection} is expected. For completeness, the evolution of the $m=1$ energies and growth rates is shown in Fig.~\ref{fig:E-g_T-1_long}(a) and (b), respectively.

An important observation is the following. While the $m=1$ mode was originally perturbed with $\vartheta_{0,m=1} = 0$ (i.e., core motion in the positive $x$ direction), the kink flow continuously changes its direction, as is obvious from the arrows drawn along with the $\phi$ contours in Figs.~\ref{fig:snaps_3tm_T-1a} and \ref{fig:snaps_3tm_T-1b}. Note that the core does not rotate; merely the direction of its translational motion alters (in RMHD, $\phi_{0,0} = 0$ at all times if it is zero initially). The core's motion is quantified and shown in more detail in Fig.~\ref{fig:E-g_T-1_long}(c), using the angles $\vartheta^{\rm kink}_{\rm mag}$ and $\vartheta^{\rm kink}_{\rm kin}$. These are defined as
\begin{subequations}
\begin{align}
\vartheta^{\rm kink}_{\rm mag} &= \tan^{-1}\left[\frac{ \int{\rm d}r.r {\rm Im}\{\psi_{1,1}(r)\} }{ \int{\rm d}r.r {\rm Re}\{\psi_{1,1}(r)\} } \right]
\\
\vartheta^{\rm kink}_{\rm kin} &= -\tan^{-1}\left[\frac{ \int{\rm d}r.r {\rm Re}\{\phi_{1,1}(r)\} }{ \int{\rm d}r.r {\rm Im}\{\phi_{1,1}(r)\} } \right],
\end{align}
\label{eq:ph-kink}
\end{subequations}

\noindent (integration interval: $0 \leq r \leq 0.07$) and give an approximate image of the core's motion when the kink amplitude is not too large. During the linear phase (i) both angles are zero, in agreement with the initial perturbation. During the driving phase (ii) the angles switch to $\vartheta^{\rm kink}_{\rm mag} \approx \vartheta^{\rm kink}_{\rm kin} \approx 3^\circ$ (determined by the driving modes). After the transition to turbulence in phase (iii), the direction of the kink flow varies frequently, so that the core's motion becomes rather complicated.

%-------------------------------------------
\subsection{Case (T-2): Kink saturation and partial reconnection}

Here we discuss the long-term evolution in Case (T-2). Snapshots are presented in Figs.~\ref{fig:snaps_3tm_T-2a} and \ref{fig:snaps_3tm_T-2b}. In Fig.~\ref{fig:E-g_T-2_long} the evolution of (a) the $m=1$ energies, (b) corresponding growth rates, and (c) the core's location and kink flow direction are shown. Note that the early evolution is very similar to that of Case (T-1) described in Section \ref{sec:kink}: It consists of (i) linear growth, (ii) nonlinearly driven fast growth, and (iii) transition to the turbulent regime with gradually decreasing growth rates (here $Pr = 0.01$). We may thus omit the further discussion of these stages, referring to Sec.~\ref{sec:kink} above.

Similarly to Case (T-1), the first macroscopically observable event is an annular collapse due to high-$m$ TTMs without significant displacement of the core [Fig.~\ref{fig:snaps_3tm_T-2a}(A)]. Again, the $q(r)$ profile is flattened in the inter-resonance region. Subsequently the $m=1$ mode grows inside the turbulent belt [Fig.~\ref{fig:snaps_3tm_T-2a}(B)]. However, in contrast to Case (T-1), here the $m=1$ mode saturates after reaching a relatively large amplitude [Fig.~\ref{fig:snaps_3tm_T-2a}(C)]. This occurs at $t=962$, as can clearly be seen in the $m=1$ magnetic energy, $\EMag_{1,1}$, and the associated growth rate, $\gMag_{1,1}$ [Fig.~\ref{fig:E-g_T-2_long}(a) and (b)]. The direction of the kink flow reverses, as is obvious from Snapshots~(C)--(E) (Figs.~\ref{fig:snaps_3tm_T-2a} and \ref{fig:snaps_3tm_T-2b}) and from the $180$-degree jump in $\vartheta^{\rm kink}_{\rm kin}$ in Fig.~\ref{fig:E-g_T-2_long}(c). Afterwards, the kink amplitude decays [Fig.~\ref{fig:snaps_3tm_T-2b}(E)], overshoots, and grows again in a different direction [Fig.~\ref{fig:snaps_3tm_T-2b}(F)].

The saturation of the $m=1$ mode observed here seems to be due to an island-like structure developing ``in front'' of the displaced core, when the latter approaches the periphery. The island remains trapped, i.e., it is not expelled in the poloidal direction. Consequently core reconnection takes place at two separate points which in the present case are located an angle $\Delta\vartheta \approx 116^o$ apart, as indicated in Fig.~\ref{fig:snaps_3tm_T-2a}(C). Since each reconnected flux surface adds to the island's width, this structure can counter the internal kink, induce a rebound and send the core back into the center. This scenario is realized in the present case.

%-------------------------------------------
\subsection{Modulation of the kink flow}

In addition to the kink flow changing its direction, at certain times we observe an $m=1$ modulation of the $\phi$ contours in the core's interior. In Case (T-1) this occurs around $t=400$, Snapshot~(D) [Fig.~\ref{fig:snaps_3tm_T-1b}]. This modulation of the $\ExB$ drift velocity is not strong enough to visibly alter the $\psi$ contours in the core, which therefore remain circular (not shown). Further details can be seen in Fig.~\ref{fig:vr_T-1}, where the profile of the radial velocity $v_r \propto \phi/r$ is shown at several times in the interval $360 \leq t \leq 510$. Some peaks in $v_r$, like those near the magnetic axis, perform oscillations in the manner of a standing wave [left-hand side in Fig.~\ref{fig:vr_T-1}(b)]. Other peaks, like those in the region $0.2 \lesssim r \lesssim 0.35$, do not change their signs until about $t \approx 500$ [Fig.~\ref{fig:vr_T-1}(c)]. The radial wavelength of the modulation is observed to change relatively slowly. Typically, it measures between 1/2 and 1/5 of the core's radius. Realizations of this $m=1$ modulation can also be observed in Case (T-2) (Figs.~\ref{fig:snaps_3tm_T-2a} and \ref{fig:snaps_3tm_T-2b}).

%===========================================
\section{Discussion and Conclusions}
\label{sec:conclusion}

We have studied the nonlinear evolution of the $m=1$ internal kink mode in a configuration with three $q=1$ resonant surfaces where high-$m$ TTMs are strongly unstable and lead to an annular collapse. The latter leaves behind a turbulent belt around the unreconnected core plasma. The simulation results show that high-$m$ TTMs and MHD turbulence, which are localized in an annular region, are able to strongly affect the evolution of the $m=1$ internal kink mode, which is a global instability. We conclude that multiple tearing modes (here, TTMs) and MHD turbulence may play a significant role during partial, compound or full sawtooth crashes in tokamak plasmas, as will be discussed in the following.

In the beginning, a fast sawtooth trigger, defined as a sudden transition from slow to rapid growth, was realized: after a phase of slow linear growth, rapidly growing $q=1$ TTMs give rise to a new fast growing nonlinear $m=1$ mode. This instability reaches an observable amplitude within a time much shorter than expected from the linear growth rate. Moreover, the transition to the fully nonlinear (turbulent) regime occurs via a phase of explosive growth when the Prandtl number is large ($Pr \gtrsim 10$). As proposed in Ref.~\cite{Bierwage05a} this enhancement of the $m=1$ mode due to high-$m$ TTMs is a possible mechanism for the fast sawtooth trigger, provided that multiple $q=1$ resonant surfaces are formed during the sawtooth ramp.
%Let us note that there is evidence that the dominance of high-$m$ TTMs is independent of the reconnection mechanism. Preliminary results show that the dispersion relations of TTMs destabilized by electron inertia seem to have similar properties as for resistive modes, especially with regard to the large value of $\mPeak$.

During the further evolution, the turbulence in the collapsed annular region was seen to alter the direction of the kink flow responsible for the core displacement. Both continuing growth [Case (T-1)] and saturation of the kink [Case (T-2)] were observed, which shows that full as well as partial reconnection may occur in this setting. It was also found that the effect of the turbulence was not limited to the collapsed annular region and the overall motion of the core inside this turbulent belt: Perturbations of the electrostatic potential were even found to penetrate into the central core region in the form of an $m=1$ modulation on top of the kink flow.

The converse effect, i.e., the influence of the core displacement on the surrounding turbulence has not been addressed and is left for future study. This is expected to be important since (a) the core displacement changes the geometry of the turbulent region and (b) the return flows of the internal kink are likely to interact with the turbulence. One particular question to be addressed is whether and how the core contributes to the formation of the trapped island that prevents further reconnection in Case (T-2) [cf.~Fig.~\ref{fig:snaps_3tm_T-2a}(C)].

Our results agree with some aspects of the partial sawtooth crash scenario suggested in earlier studies: A ``shoulder'' on the $q(r)$ profile forms where $q \simeq 1$, and this region is governed by electromagnetic turbulence (e.g., \cite{Rogister89, Rogister90, Porcelli96}). Indeed, the conjectures made by these authors imply that continued growth of the $m=1$ mode [as in Case (T-1)] must be prevented by some means, so that the partial collapse remains partial. We have demonstrated that MHD turbulence is one possible mechanism leading to a saturation of the internal kink [as in Case (T-2)]. The residual core displacement may then account for the post-cursor oscillation observed experimentally after partial reconnection events (e.g., \cite{Westerhof89, Koslowski97}).

The fact that the long-term calculations in the two cases considered here have different outcomes, namely full reconnection in Case (T-1) and partial reconnection in Case (T-2), requires commenting on. Note that the initial conditions do not differ largely, except for the linear growth rates in Case (T-1) being twice as high as in Case (T-2) due to a higher magnetic shear (cf.~Figs.~\ref{fig:spec_3tm_T-1} and \ref{fig:spec_3tm_T-2}). While we were able to demonstrate that MHD turbulence provides prerequisites for a saturation of the internal kink, the ultimate goal of determining quantitative criteria for partial reconnection requires further investigations using more realistic physical models. As mentioned in the introduction, potentially important effects to be considered include two-fluid, curvature, and finite-beta effects. In particular, if some of these would be found to have a stabilizing influence on the high-$m$ modes, which are essential in the present work, the results may be altered. To our knowledge, such studies have not been conducted for a comparable scenario, i.e., strongly coupled multiple tearing modes such as DTMs or TTMs.

In this study, MHD turbulence was generated through current-driven resistive TTMs, requiring multiple $q=1$ resonant surfaces. The inclusion of a collisionless reconnection mechanism is expected to yield higher kinetic energies \cite{BiskampDrake94} and thus stronger turbulence interacting with the internal kink. Furthermore, we conjecture that our principal result, namely that the internal kink mode can be strongly affected by tearing-mode-driven MHD turbulence, will also apply when the latter is generated by some other means, such as pressure-driven MHD or micro-instabilities \cite{Thyagaraja05}.

Let us note that there are several other mechanisms which were proposed as possible explanations for the rapid collapse and partial crash phenomena, which were discovered assuming different pre-crash conditions (e.g.,~Ref.~\cite{Hastie98} and references therein). However, the currently available experimental data is not yet conclusive enough to rule out one proposed sawtooth crash scenario or another. Moreover, the detailed evolution may vary between different machines, shots and even between sawtooth crashes of a single discharge.

Through recent progress in plasma diagnostics it has become possible to detect high-$m$ magnetic islands associated with low-order $q = m/n$ resonant surfaces \cite{Donne05}. Thus, experimental checks of our simulation results seem feasible in the near future.

%-------------------------------------------
\begin{acknowledgments}
A.B. would like to thank Y. Kishimoto, Y. Nakamura and F. Sano for valuable discussions. He also acknowledges the Max-Planck-Institut f\"{u}r Plasmaphysik in Garching for its hospitality and for providing computational resources for numerical checks. S.B. acknowledges the Graduate School of Energy Science at Kyoto University and the Center for Atomic and Molecular Technologies at Osaka University for their support and hospitality. This work was partially supported by the 21st Century COE Program at Kyoto University.
\end{acknowledgments}

%===========================================
%\bibliography{p03_ttm-nlin}

\begin{thebibliography}{37}
\expandafter\ifx\csname natexlab\endcsname\relax\def\natexlab#1{#1}\fi
\expandafter\ifx\csname bibnamefont\endcsname\relax
  \def\bibnamefont#1{#1}\fi
\expandafter\ifx\csname bibfnamefont\endcsname\relax
  \def\bibfnamefont#1{#1}\fi
\expandafter\ifx\csname citenamefont\endcsname\relax
  \def\citenamefont#1{#1}\fi
\expandafter\ifx\csname url\endcsname\relax
  \def\url#1{\texttt{#1}}\fi
\expandafter\ifx\csname urlprefix\endcsname\relax\def\urlprefix{URL }\fi
\providecommand{\bibinfo}[2]{#2}
\providecommand{\eprint}[2][]{\url{#2}}

\bibitem[{\citenamefont{Bierwage
  et~al.}(2005{\natexlab{a}})\citenamefont{Bierwage, Hamaguchi, Wakatani,
  Benkadda, and Leoncini}}]{Bierwage05a}
\bibinfo{author}{\bibfnamefont{A.}~\bibnamefont{Bierwage}},
  \bibinfo{author}{\bibfnamefont{S.}~\bibnamefont{Hamaguchi}},
  \bibinfo{author}{\bibfnamefont{M.}~\bibnamefont{Wakatani}},
  \bibinfo{author}{\bibfnamefont{S.}~\bibnamefont{Benkadda}}, \bibnamefont{and}
  \bibinfo{author}{\bibfnamefont{X.}~\bibnamefont{Leoncini}},
  \bibinfo{journal}{Phys. Rev. Lett.} \textbf{\bibinfo{volume}{94}},
  \bibinfo{pages}{065001} (\bibinfo{year}{2005}{\natexlab{a}}).

\bibitem[{\citenamefont{Bierwage
  et~al.}(2005{\natexlab{b}})\citenamefont{Bierwage, Benkadda, Hamaguchi, and
  Wakatani}}]{Bierwage05b}
\bibinfo{author}{\bibfnamefont{A.}~\bibnamefont{Bierwage}},
  \bibinfo{author}{\bibfnamefont{S.}~\bibnamefont{Benkadda}},
  \bibinfo{author}{\bibfnamefont{S.}~\bibnamefont{Hamaguchi}},
  \bibnamefont{and} \bibinfo{author}{\bibfnamefont{M.}~\bibnamefont{Wakatani}},
  \bibinfo{journal}{Phys. Plasmas} \textbf{\bibinfo{volume}{12}},
  \bibinfo{pages}{082504} (\bibinfo{year}{2005}{\natexlab{b}}).

\bibitem[{\citenamefont{Kuvshinov and Savrukhin}(1990)}]{Kuvshinov90}
\bibinfo{author}{\bibfnamefont{B.~N.} \bibnamefont{Kuvshinov}}
  \bibnamefont{and} \bibinfo{author}{\bibfnamefont{P.~V.}
  \bibnamefont{Savrukhin}}, \bibinfo{journal}{Sov. J. Plasma Phys.}
  \textbf{\bibinfo{volume}{16}}, \bibinfo{pages}{353} (\bibinfo{year}{1990}).

\bibitem[{\citenamefont{Migliuolo}(1993)}]{Migliuolo93}
\bibinfo{author}{\bibfnamefont{S.}~\bibnamefont{Migliuolo}},
  \bibinfo{journal}{Nucl. Fusion} \textbf{\bibinfo{volume}{33}},
  \bibinfo{pages}{1721} (\bibinfo{year}{1993}).

\bibitem[{\citenamefont{Hastie}(1998)}]{Hastie98}
\bibinfo{author}{\bibfnamefont{R.~J.} \bibnamefont{Hastie}},
  \bibinfo{journal}{Astrophys. Space Sci.} \textbf{\bibinfo{volume}{256}},
  \bibinfo{pages}{177} (\bibinfo{year}{1998}).

\bibitem[{\citenamefont{Porcelli et~al.}(2004)\citenamefont{Porcelli,
  Annibaldi, Borgogno, Buratti, Califano, Coelho, Giovannozzi, Grasso, Lazzaro,
  Pegoraro et~al.}}]{Porcelli04}
\bibinfo{author}{\bibfnamefont{F.}~\bibnamefont{Porcelli}},
  \bibinfo{author}{\bibfnamefont{S.}~\bibnamefont{Annibaldi}},
  \bibinfo{author}{\bibfnamefont{D.}~\bibnamefont{Borgogno}},
  \bibinfo{author}{\bibfnamefont{P.}~\bibnamefont{Buratti}},
  \bibinfo{author}{\bibfnamefont{F.}~\bibnamefont{Califano}},
  \bibinfo{author}{\bibfnamefont{R.}~\bibnamefont{Coelho}},
  \bibinfo{author}{\bibfnamefont{E.}~\bibnamefont{Giovannozzi}},
  \bibinfo{author}{\bibfnamefont{D.}~\bibnamefont{Grasso}},
  \bibinfo{author}{\bibfnamefont{E.}~\bibnamefont{Lazzaro}},
  \bibinfo{author}{\bibfnamefont{F.}~\bibnamefont{Pegoraro}},
  \bibnamefont{et~al.}, \bibinfo{journal}{Nucl. Fusion}
  \textbf{\bibinfo{volume}{44}}, \bibinfo{pages}{362} (\bibinfo{year}{2004}).

\bibitem[{\citenamefont{Soltwisch et~al.}(1987)\citenamefont{Soltwisch,
  Stodiek, Manickam, and Schl\"{u}ter}}]{Soltwisch86}
\bibinfo{author}{\bibfnamefont{H.}~\bibnamefont{Soltwisch}},
  \bibinfo{author}{\bibfnamefont{W.}~\bibnamefont{Stodiek}},
  \bibinfo{author}{\bibfnamefont{J.}~\bibnamefont{Manickam}}, \bibnamefont{and}
  \bibinfo{author}{\bibfnamefont{J.}~\bibnamefont{Schl\"{u}ter}}, in
  \emph{\bibinfo{booktitle}{Plasma Physics and Controlled Nuclear Fusion
  Research 1986}} (\bibinfo{publisher}{International Atomic Energy Agency},
  \bibinfo{address}{Vienna}, \bibinfo{year}{1987}), vol.~\bibinfo{volume}{1},
  p. \bibinfo{pages}{263}, \bibinfo{note}{{IAEA-CN-47/A-V-1}}.

\bibitem[{\citenamefont{Edwards et~al.}(1986)\citenamefont{Edwards, Campbell,
  Engelhardt, Fahrbach, Gill, Granetz, Tsuji, Tubbing, Weller, Wesson
  et~al.}}]{Edwards86}
\bibinfo{author}{\bibfnamefont{A.~W.} \bibnamefont{Edwards}},
  \bibinfo{author}{\bibfnamefont{D.~J.} \bibnamefont{Campbell}},
  \bibinfo{author}{\bibfnamefont{W.~W.} \bibnamefont{Engelhardt}},
  \bibinfo{author}{\bibfnamefont{H.-U.} \bibnamefont{Fahrbach}},
  \bibinfo{author}{\bibfnamefont{R.~D.} \bibnamefont{Gill}},
  \bibinfo{author}{\bibfnamefont{R.~S.} \bibnamefont{Granetz}},
  \bibinfo{author}{\bibfnamefont{S.}~\bibnamefont{Tsuji}},
  \bibinfo{author}{\bibfnamefont{B.~J.~D.} \bibnamefont{Tubbing}},
  \bibinfo{author}{\bibfnamefont{A.}~\bibnamefont{Weller}},
  \bibinfo{author}{\bibfnamefont{J.}~\bibnamefont{Wesson}},
  \bibnamefont{et~al.}, \bibinfo{journal}{Phys. Rev. Lett.}
  \textbf{\bibinfo{volume}{57}}, \bibinfo{pages}{210} (\bibinfo{year}{1986}).

\bibitem[{\citenamefont{Levinton et~al.}(1993)\citenamefont{Levinton, Batha,
  Yamada, and Zarnstorff}}]{Levinton93}
\bibinfo{author}{\bibfnamefont{F.~M.} \bibnamefont{Levinton}},
  \bibinfo{author}{\bibfnamefont{S.~H.} \bibnamefont{Batha}},
  \bibinfo{author}{\bibfnamefont{S.~H.} \bibnamefont{Yamada}},
  \bibnamefont{and} \bibinfo{author}{\bibfnamefont{M.~C.}
  \bibnamefont{Zarnstorff}}, \bibinfo{journal}{Phys. Fluids B}
  \textbf{\bibinfo{volume}{5}}, \bibinfo{pages}{2554} (\bibinfo{year}{1993}).

\bibitem[{\citenamefont{Aydemir}(1992)}]{Aydemir92}
\bibinfo{author}{\bibfnamefont{A.~Y.} \bibnamefont{Aydemir}},
  \bibinfo{journal}{Phys. Fluids B} \textbf{\bibinfo{volume}{4}},
  \bibinfo{pages}{3469} (\bibinfo{year}{1992}).

\bibitem[{\citenamefont{Wang and Bhattacharjee}(1993)}]{WangBhatt93}
\bibinfo{author}{\bibfnamefont{X.}~\bibnamefont{Wang}} \bibnamefont{and}
  \bibinfo{author}{\bibfnamefont{A.}~\bibnamefont{Bhattacharjee}},
  \bibinfo{journal}{Phys. Rev. Lett.} \textbf{\bibinfo{volume}{70}},
  \bibinfo{pages}{1627} (\bibinfo{year}{1993}).

\bibitem[{\citenamefont{Wang and Bhattacharjee}(1995)}]{WangBhatt95}
\bibinfo{author}{\bibfnamefont{X.}~\bibnamefont{Wang}} \bibnamefont{and}
  \bibinfo{author}{\bibfnamefont{A.}~\bibnamefont{Bhattacharjee}},
  \bibinfo{journal}{Phys. Plasmas} \textbf{\bibinfo{volume}{2}},
  \bibinfo{pages}{171} (\bibinfo{year}{1995}).

\bibitem[{\citenamefont{Biskamp and Sato}(1997)}]{BiskampSato97}
\bibinfo{author}{\bibfnamefont{D.}~\bibnamefont{Biskamp}} \bibnamefont{and}
  \bibinfo{author}{\bibfnamefont{T.}~\bibnamefont{Sato}},
  \bibinfo{journal}{Phys. Plasmas} \textbf{\bibinfo{volume}{4}},
  \bibinfo{pages}{1326} (\bibinfo{year}{1997}).

\bibitem[{\citenamefont{Sundaram and Sen}(1980)}]{Sundaram80}
\bibinfo{author}{\bibfnamefont{A.~K.} \bibnamefont{Sundaram}} \bibnamefont{and}
  \bibinfo{author}{\bibfnamefont{A.}~\bibnamefont{Sen}},
  \bibinfo{journal}{Phys. Rev. Lett.} \textbf{\bibinfo{volume}{44}},
  \bibinfo{pages}{322} (\bibinfo{year}{1980}).

\bibitem[{\citenamefont{Sen and Sundaram}(1981)}]{Sen81}
\bibinfo{author}{\bibfnamefont{A.}~\bibnamefont{Sen}} \bibnamefont{and}
  \bibinfo{author}{\bibfnamefont{A.~K.} \bibnamefont{Sundaram}},
  \bibinfo{journal}{Phys. Fluids} \textbf{\bibinfo{volume}{24}},
  \bibinfo{pages}{1303} (\bibinfo{year}{1981}).

\bibitem[{\citenamefont{Thyagaraja et~al.}(1992)\citenamefont{Thyagaraja,
  Hazeltine, and Aydemir}}]{Thyagaraja92}
\bibinfo{author}{\bibfnamefont{A.}~\bibnamefont{Thyagaraja}},
  \bibinfo{author}{\bibfnamefont{R.~D.} \bibnamefont{Hazeltine}},
  \bibnamefont{and} \bibinfo{author}{\bibfnamefont{A.~Y.}
  \bibnamefont{Aydemir}}, \bibinfo{journal}{Phys. Fluids B}
  \textbf{\bibinfo{volume}{4}}, \bibinfo{pages}{2733} (\bibinfo{year}{1992}).

\bibitem[{\citenamefont{Thyagaraja and Haas}(1993)}]{Thyagaraja93}
\bibinfo{author}{\bibfnamefont{A.}~\bibnamefont{Thyagaraja}} \bibnamefont{and}
  \bibinfo{author}{\bibfnamefont{F.~A.} \bibnamefont{Haas}},
  \bibinfo{journal}{Phys. Fluids B} \textbf{\bibinfo{volume}{5}},
  \bibinfo{pages}{3252} (\bibinfo{year}{1993}).

\bibitem[{\citenamefont{Matthaeus and Lamkin}(1986)}]{Matthaeus86}
\bibinfo{author}{\bibfnamefont{W.~H.} \bibnamefont{Matthaeus}}
  \bibnamefont{and} \bibinfo{author}{\bibfnamefont{S.~L.}
  \bibnamefont{Lamkin}}, \bibinfo{journal}{Phys. Fluids}
  \textbf{\bibinfo{volume}{29}}, \bibinfo{pages}{2513} (\bibinfo{year}{1986}).

\bibitem[{\citenamefont{Drake et~al.}(1994)\citenamefont{Drake, Kleva, and
  Mandt}}]{Drake94}
\bibinfo{author}{\bibfnamefont{J.~F.} \bibnamefont{Drake}},
  \bibinfo{author}{\bibfnamefont{R.~G.} \bibnamefont{Kleva}}, \bibnamefont{and}
  \bibinfo{author}{\bibfnamefont{M.~E.} \bibnamefont{Mandt}},
  \bibinfo{journal}{Phys. Rev. Lett.} \textbf{\bibinfo{volume}{73}},
  \bibinfo{pages}{1251} (\bibinfo{year}{1994}).

\bibitem[{\citenamefont{Kim and Diamond}(2001)}]{Kim01}
\bibinfo{author}{\bibfnamefont{E.-J.} \bibnamefont{Kim}} \bibnamefont{and}
  \bibinfo{author}{\bibfnamefont{P.~H.} \bibnamefont{Diamond}},
  \bibinfo{journal}{Astrophys. J.} \textbf{\bibinfo{volume}{556}},
  \bibinfo{pages}{1052} (\bibinfo{year}{2001}).

\bibitem[{\citenamefont{Aparicio et~al.}(1998)\citenamefont{Aparicio, Haines,
  Hastie, and Wainwright}}]{Aparicio98}
\bibinfo{author}{\bibfnamefont{J.}~\bibnamefont{Aparicio}},
  \bibinfo{author}{\bibfnamefont{M.~G.} \bibnamefont{Haines}},
  \bibinfo{author}{\bibfnamefont{R.~J.} \bibnamefont{Hastie}},
  \bibnamefont{and} \bibinfo{author}{\bibfnamefont{J.~P.}
  \bibnamefont{Wainwright}}, \bibinfo{journal}{Phys. Plasmas}
  \textbf{\bibinfo{volume}{5}}, \bibinfo{pages}{3180} (\bibinfo{year}{1998}).

\bibitem[{\citenamefont{Meiss et~al.}(1982)\citenamefont{Meiss, Hazeltine,
  Diamond, and Mahajan}}]{Meiss82}
\bibinfo{author}{\bibfnamefont{J.~D.} \bibnamefont{Meiss}},
  \bibinfo{author}{\bibfnamefont{R.~D.} \bibnamefont{Hazeltine}},
  \bibinfo{author}{\bibfnamefont{P.~H.} \bibnamefont{Diamond}},
  \bibnamefont{and} \bibinfo{author}{\bibfnamefont{S.~M.}
  \bibnamefont{Mahajan}}, \bibinfo{journal}{Phys. Fluids}
  \textbf{\bibinfo{volume}{25}}, \bibinfo{pages}{815} (\bibinfo{year}{1982}).

\bibitem[{\citenamefont{Kaw et~al.}(1979)\citenamefont{Kaw, Valeo, and
  Rutherford}}]{Kaw79}
\bibinfo{author}{\bibfnamefont{P.~K.} \bibnamefont{Kaw}},
  \bibinfo{author}{\bibfnamefont{E.~J.} \bibnamefont{Valeo}}, \bibnamefont{and}
  \bibinfo{author}{\bibfnamefont{P.~H.} \bibnamefont{Rutherford}},
  \bibinfo{journal}{Phys. Rev. Lett.} \textbf{\bibinfo{volume}{43}},
  \bibinfo{pages}{1398} (\bibinfo{year}{1979}).

\bibitem[{\citenamefont{Lazarian and Vishniac}(1999)}]{Lazarian99}
\bibinfo{author}{\bibfnamefont{A.}~\bibnamefont{Lazarian}} \bibnamefont{and}
  \bibinfo{author}{\bibfnamefont{E.~T.} \bibnamefont{Vishniac}},
  \bibinfo{journal}{Astrophys. J.} \textbf{\bibinfo{volume}{517}},
  \bibinfo{pages}{700} (\bibinfo{year}{1999}).

\bibitem[{\citenamefont{Dubois and Samain}(1980)}]{Dubois80}
\bibinfo{author}{\bibfnamefont{M.}~\bibnamefont{Dubois}} \bibnamefont{and}
  \bibinfo{author}{\bibfnamefont{A.}~\bibnamefont{Samain}},
  \bibinfo{journal}{Nucl. Fusion} \textbf{\bibinfo{volume}{20}},
  \bibinfo{pages}{1101} (\bibinfo{year}{1980}).

\bibitem[{\citenamefont{Bussac and Pellat}(1987)}]{Bussac87}
\bibinfo{author}{\bibfnamefont{M.~N.} \bibnamefont{Bussac}} \bibnamefont{and}
  \bibinfo{author}{\bibfnamefont{R.}~\bibnamefont{Pellat}},
  \bibinfo{journal}{Phys. Rev. Lett.} \textbf{\bibinfo{volume}{59}},
  \bibinfo{pages}{2650} (\bibinfo{year}{1987}).

\bibitem[{\citenamefont{Nishimura et~al.}(1999)\citenamefont{Nishimura, Callen,
  and Hegna}}]{Nishimura99}
\bibinfo{author}{\bibfnamefont{Y.}~\bibnamefont{Nishimura}},
  \bibinfo{author}{\bibfnamefont{J.~D.} \bibnamefont{Callen}},
  \bibnamefont{and} \bibinfo{author}{\bibfnamefont{C.~C.} \bibnamefont{Hegna}},
  \bibinfo{journal}{Phys. Plasmas} \textbf{\bibinfo{volume}{6}},
  \bibinfo{pages}{4685} (\bibinfo{year}{1999}).

\bibitem[{\citenamefont{Porcelli et~al.}(1996)\citenamefont{Porcelli, Boucher,
  and Rosenbluth}}]{Porcelli96}
\bibinfo{author}{\bibfnamefont{F.}~\bibnamefont{Porcelli}},
  \bibinfo{author}{\bibfnamefont{D.}~\bibnamefont{Boucher}}, \bibnamefont{and}
  \bibinfo{author}{\bibfnamefont{M.~N.} \bibnamefont{Rosenbluth}},
  \bibinfo{journal}{Plasma Phys. Control. Fusion}
  \textbf{\bibinfo{volume}{38}}, \bibinfo{pages}{2163} (\bibinfo{year}{1996}).

\bibitem[{\citenamefont{Strauss}(1976)}]{Strauss76}
\bibinfo{author}{\bibfnamefont{H.~R.} \bibnamefont{Strauss}},
  \bibinfo{journal}{Phys. Fluids} \textbf{\bibinfo{volume}{19}},
  \bibinfo{pages}{134} (\bibinfo{year}{1976}).

\bibitem[{\citenamefont{Nishikawa and Wakatani}(2000)}]{NishikawaWakatani}
\bibinfo{author}{\bibfnamefont{K.}~\bibnamefont{Nishikawa}} \bibnamefont{and}
  \bibinfo{author}{\bibfnamefont{M.}~\bibnamefont{Wakatani}},
  \emph{\bibinfo{title}{Plasma Physics}} (\bibinfo{publisher}{Springer,
  Berlin}, \bibinfo{year}{2000}).

\bibitem[{\citenamefont{Rogister et~al.}(1989)\citenamefont{Rogister, Singh,
  and Kaleck}}]{Rogister89}
\bibinfo{author}{\bibfnamefont{A.~L.~M.} \bibnamefont{Rogister}},
  \bibinfo{author}{\bibfnamefont{R.}~\bibnamefont{Singh}}, \bibnamefont{and}
  \bibinfo{author}{\bibfnamefont{A.}~\bibnamefont{Kaleck}},
  \bibinfo{journal}{Nucl. Fusion} \textbf{\bibinfo{volume}{29}},
  \bibinfo{pages}{1175} (\bibinfo{year}{1989}).

\bibitem[{\citenamefont{Rogister et~al.}(1990)\citenamefont{Rogister, Kaleck,
  Psimopoulos, and Hasselberg}}]{Rogister90}
\bibinfo{author}{\bibfnamefont{A.}~\bibnamefont{Rogister}},
  \bibinfo{author}{\bibfnamefont{A.}~\bibnamefont{Kaleck}},
  \bibinfo{author}{\bibfnamefont{M.}~\bibnamefont{Psimopoulos}},
  \bibnamefont{and}
  \bibinfo{author}{\bibfnamefont{G.}~\bibnamefont{Hasselberg}},
  \bibinfo{journal}{Phys. Fluids B} \textbf{\bibinfo{volume}{2}},
  \bibinfo{pages}{953} (\bibinfo{year}{1990}).

\bibitem[{\citenamefont{Westerhof et~al.}(1989)\citenamefont{Westerhof,
  Smeulders, and Cardozo}}]{Westerhof89}
\bibinfo{author}{\bibfnamefont{E.}~\bibnamefont{Westerhof}},
  \bibinfo{author}{\bibfnamefont{P.}~\bibnamefont{Smeulders}},
  \bibnamefont{and} \bibinfo{author}{\bibfnamefont{N.~L.}
  \bibnamefont{Cardozo}}, \bibinfo{journal}{Nucl. Fusion}
  \textbf{\bibinfo{volume}{29}}, \bibinfo{pages}{1056} (\bibinfo{year}{1989}).

\bibitem[{\citenamefont{Koslowski et~al.}(1997)\citenamefont{Koslowski, Fuchs,
  Kr\"{a}mer-Flecken, Rapp, and {the TEXTOR-94 team}}}]{Koslowski97}
\bibinfo{author}{\bibfnamefont{H.~R.} \bibnamefont{Koslowski}},
  \bibinfo{author}{\bibfnamefont{G.}~\bibnamefont{Fuchs}},
  \bibinfo{author}{\bibfnamefont{A.}~\bibnamefont{Kr\"{a}mer-Flecken}},
  \bibinfo{author}{\bibfnamefont{J.}~\bibnamefont{Rapp}}, \bibnamefont{and}
  \bibinfo{author}{\bibnamefont{{the TEXTOR-94 team}}},
  \bibinfo{journal}{Plasma Phys. Control. Fusion}
  \textbf{\bibinfo{volume}{39}}, \bibinfo{pages}{B325} (\bibinfo{year}{1997}).

\bibitem[{\citenamefont{Biskamp and Drake}(1994)}]{BiskampDrake94}
\bibinfo{author}{\bibfnamefont{D.}~\bibnamefont{Biskamp}} \bibnamefont{and}
  \bibinfo{author}{\bibfnamefont{J.~F.} \bibnamefont{Drake}},
  \bibinfo{journal}{Phys. Rev. Lett.} \textbf{\bibinfo{volume}{73}},
  \bibinfo{pages}{971} (\bibinfo{year}{1994}).

\bibitem[{\citenamefont{Thyagaraja et~al.}(2005)\citenamefont{Thyagaraja,
  Knight, de~Baar, Hogeweij, and Min}}]{Thyagaraja05}
\bibinfo{author}{\bibfnamefont{A.}~\bibnamefont{Thyagaraja}},
  \bibinfo{author}{\bibfnamefont{P.~J.} \bibnamefont{Knight}},
  \bibinfo{author}{\bibfnamefont{M.~R.} \bibnamefont{de~Baar}},
  \bibinfo{author}{\bibfnamefont{G.~M.~D.} \bibnamefont{Hogeweij}},
  \bibnamefont{and} \bibinfo{author}{\bibfnamefont{E.}~\bibnamefont{Min}},
  \bibinfo{journal}{Phys. Plasmas} \textbf{\bibinfo{volume}{12}},
  \bibinfo{pages}{090907} (\bibinfo{year}{2005}).

\bibitem[{\citenamefont{Donn\'{e} et~al.}(2005)\citenamefont{Donn\'{e}, van
  Gorkom, Udintsev, Domier, Kr\"{a}mer-Flecken, {Luhmann, Jr.}, Sch\"{u}ller,
  and {TEXTOR team}}}]{Donne05}
\bibinfo{author}{\bibfnamefont{A.~J.~H.} \bibnamefont{Donn\'{e}}},
  \bibinfo{author}{\bibfnamefont{J.~C.} \bibnamefont{van Gorkom}},
  \bibinfo{author}{\bibfnamefont{V.~S.} \bibnamefont{Udintsev}},
  \bibinfo{author}{\bibfnamefont{C.~W.} \bibnamefont{Domier}},
  \bibinfo{author}{\bibfnamefont{A.}~\bibnamefont{Kr\"{a}mer-Flecken}},
  \bibinfo{author}{\bibfnamefont{N.~C.} \bibnamefont{{Luhmann, Jr.}}},
  \bibinfo{author}{\bibfnamefont{F.~C.} \bibnamefont{Sch\"{u}ller}},
  \bibnamefont{and} \bibinfo{author}{\bibnamefont{{TEXTOR team}}},
  \bibinfo{journal}{Phys. Rev. Lett.} \textbf{\bibinfo{volume}{94}},
  \bibinfo{pages}{085001} (\bibinfo{year}{2005}).

\end{thebibliography}
%\end{document}

%===========================================

\end{document}